\documentclass[11pt,preprint]{aastex}
\usepackage{color}
\usepackage{graphicx}

\newif\ifdraftmodep
\draftmodepfalse

\newcommand{\NOTE}[1]{\ifdraftmodep {\color {red} [{\it #1}]} \fi}

\newif\ifapjp
\apjpfalse
\usepackage{ifthen,color,graphicx}

\newcommand{\outlinestart}[1]
{
\ifthenelse{\boolean{#1}}{\begin{enumerate}}{}
}

\newcommand{\outline}[2]
{
\ifthenelse{\boolean{#1}}{\it \color {red} \item #2 \rm \color {black}\\}{}
}

\newcommand{\outlineend}[1]
{
\ifthenelse{\boolean{#1}}{\end{enumerate}}{}
}

\newboolean{draft}
\setboolean{draft}{false}

\begin{document}
\title{Quasar Spectrum Classification with PCA - II:\\
Introduction of Five Classes, 
Artificial Quasar Spectrum, 
the Mean Flux Correction Factor $\delta F$,
and the Identification of Emission Lines in the Ly$\alpha$ Forest
}

\author{ Nao Suzuki\altaffilmark{1}\\ 
Center for Astrophysics and Space Sciences,\\
University of California, San Diego,\\
La Jolla; CA 92093-0424
\altaffiltext{1} 
{E-mail: suzuki@genesis.ucsd.edu}
}

\begin{abstract}

We investigate the variety in quasar UV spectra 
($\lambda\lambda$1020-1600) with emphasis on the weak emission 
lines in the Ly$\alpha$ forest region
using principal component analysis (PCA).
This paper is a continuation of \citet[][Paper I]{suzuki05a}, but with
a different approach.
We use 50 smooth continuum fitted quasar spectra (0.14 $< z <$ 1.04) 
taken by the {\it Hubble Space Telescope} ({\it HST}) Faint Object 
Spectrograph.  
There are no broad absorption line quasars included in these 50 spectra.
The first, second and third principal component spectra (PCS) account
for 63.4, 14.5 and 6.2\% of the variance respectively, and
the first seven PCS take 96.1\% of the total variance.
The first PCS carries Ly$\alpha$, Ly$\beta$ and high ionization emission 
line features (\ion{O}{6}, \ion{N}{5}, Si~IV, C~IV) that are sharp and strong.
The second PCS has low ionization emission line features 
(Fe~II, Fe~III, Si~II, C~II) that are broad and rounded. 
Three weak emission lines in the Ly$\alpha$ forest are identified as
Fe~II $\lambda1070.95$, Fe~II+Fe~III $\lambda1123.17$, and 
C~III$^{*}$ $\lambda1175.88$.
Using the first two standardized PCS coefficients, we introduce five 
classifications: {\it Class Zero} and {\it Classes I-IV}. 
These classifications will guide us in finding the continuum level 
in the Ly$\alpha$ forest. 
We show that the emission lines in the Ly$\alpha$ forest 
become prominent for {\it Classes III} and {\it IV}.
By actively using PCS, we can generate artificial quasar spectra that
are useful in testing the detection of quasars, DLAs, 
and the continuum calibration.  
We provide 10,000 artificially generated spectra.
We show that the power-law extrapolated continuum 
is inadequate to perform precise measurements 
of the mean flux in the Ly$\alpha$ forest because of the weak emission 
lines and the extended tails of Ly$\alpha$ and Ly$\beta$/O~VI emission lines.
We introduce a correction factor $\delta F$ so that the true mean flux
$\langle F \rangle$ can be related to 
$\langle F_{Power-Law} \rangle $
as measured using power-law continuum extrapolation by : 
$\langle F \rangle = \langle F_{Power-Law} \rangle \cdot \delta F$.
The correction factor $\delta F$ ranges from 0.84 to 1.05 with a
mean of 0.947 and a standard deviation of 0.031 for our 50 quasars.
This result means that using power-law extrapolation
we miss 5.3\% of flux on average and we show that
there are cases where we would miss 16\% of flux.
These corrections are essential in the study of the
intergalactic medium at high redshift 
in order to achieve precise measurements of
physical properties, cosmological parameters, and 
the reionization epoch.

\end{abstract}


\section{Introduction \NOTE{section intro} } 
\label{sec:intro}
\outline{draft}{QSO Spectra and the need for objective classification}
It is important for the study of quasar absorption lines to understand
the shape of the continua in the Ly$\alpha$ forest
from which we study the physical properties of the intergalactic 
medium (IGM) and extract cosmological 
parameters \citep{kirkman03a,tytler04c}.
It is the uncertainty of continuum fitting to the Ly$\alpha$ forest
that makes precise measurements difficult\citep{croft02a,suzuki03b,jena04a}.
Thus, we wish to have a simple and objective quasar spectrum classification
scheme that enables us to describe the global shape of the continuum 
as well as the individual emission line profiles.
The principal component analysis (PCA), also known as 
Karhunen-Lo\`eve expansion, is one of the best methods to carry 
out such classification.

\outline{draft}{A brief history of PCA}
PCA enables one to summarize the information contained in a large data set,
and it is widely used in many areas of astronomy \citep[][and references 
therein]{cabanac02a}.
\citet{francis92} applied PCA to quasar spectra 
using 232 quasar spectra (1.8 $<$ z $<$ 2.2 ;  $\lambda\lambda1150-2000$) 
from the Large Bright Quasar Survey  \citep[LBQS;][]{hewett95,hewett01}.
They showed that the first three principal components account for 75\%
of the variance.
\citet{boroson92a} used 87 low redshift quasar spectra ( z $<$ 0.5)
and showed the anticorrelation between the equivalent width
of Fe~II and [O~III] and the correlation between the luminosity, 
the strength of He~II $\lambda$4686, and the slope index $\alpha_{ox}$.
\citet{boroson02a} investigated the relation between the first
two principal components and the physical properties such as
black hole mass, luminosity, and radio activity.
\citet{shang03a} studied 22 low redshift UV and optical quasar spectra 
(z $<$ 0.5 ; $\lambda\lambda1171-6607$) and
showed the relation between the first principal component and 
the Baldwin effect, which is the anticorrelation between the luminosity and 
the equivalent width of the C~IV emission line.
\citet{yip04b} applied PCA to the 16,707 Sloan Digital Sky Survey quasar 
spectra (0.08 $< z <$ 5.41 ; $\lambda\lambda900-8000$) and reported
that the spectral classification depends on the redshift and luminosity,
and that there is no compact set of eigenspectra that can describe the 
majority of variations.
They also showed the relationship between eigencoefficients and 
the Baldwin effect.

\outline{draft}{Our Previous Work}
In \citet[][hereafter Paper~I]{suzuki05a} 
we analyzed 50 continua fitted quasar spectra
taken by the Hubble Space Telescope ({\it HST}) Faint Object Spectrograph.
Since they are at low redshifts ($0.14 < z < 1.04$), 
and the Ly$\alpha$ forest lines are not so dense, 
we can see and correctly fit the continuum to the spectra.
Using PCA we attempted to predict the continuum in the Ly$\alpha$ forest 
where
the continuum levels are hard to see because of the 
abundance of the IGM absorptions.
Although we succeeded in predicting the shape of the weak emission lines
in the Ly$\alpha$ forest region, our prediction suffers
systematic errors of $3-30$\%.
This paper is a continuation of Paper~I, but with a different approach.

\outline{draft}{Goal \& Structure of This Paper}
The goal of this paper is to explore the variety of quasar UV spectra
in the following manner:
1. we clarify the PCA formulation in order to describe quasar UV spectra
quantitatively (\S \ref{sec:b}) using eigenspectra or the principal
component spectra (PCS),
2. we introduce five classes of quasar UV spectra to help us understand
the variety of quasar spectra qualitatively (\S \ref{sec:i}),
3. we introduce the idea of artificial quasar spectra (\S \ref{sec:f})
and 4. the mean flux correction factor $\delta F$ 
(\S \ref{sec:g}), and 
5. we report the identities of three weak emission lines in the 
Ly$\alpha$ forest (Appendix).

\section{Data \NOTE{section a} } \label{sec:a}
\outline{draft}{Bechtold 334 HST FOS spectra, continua fitted, BAL removed,}
We use the same 50 {\it HST} FOS spectra from Paper I, and the detailed
description is therein.
Here we summarize the 50 {\it HST} spectra.
These 50 quasar spectra are a subset of the 334 high resolution {\it HST} 
FOS spectra (R $\sim$ 1300) collected and calibrated by \citet{bechtold02} 
which include
all of the {\it HST} QSO Absorption Line Key Project's data 
\citep{bahcall93,bahcall96,jannuzi96}.
\citet{bechtold02} identified both intergalactic and interstellar medium 
lines and corrected for Galactic extinction.

\outline{draft}{Suzuki Correction}
For each spectrum, we combined the individual exposures and remeasured the 
redshift using the peak of the Ly$\alpha$ emission line.
Then we brought the spectrum to the rest frame and rebinned it
into 0.5 \AA\ pixels.
We masked the identified absorption lines and fitted a smooth Chebyshev
polynomial curve.
We normalized the flux by taking the average of 21 pixels 
around 1280 \AA\ where no emission line is seen.

\outline{draft}{Suzuki Selection}
We chose the wavelength range from 1020 to 1600 \AA\ so that we 
can cover the Ly$\alpha$ forest between the Ly$\beta$ and 
the Ly$\alpha$ emission lines while maximizing the number of spectra. 
We removed broad absorption line (BAL) quasars from the sample since we are 
interested in emission line profiles and the continuum shape.
Lastly, we removed quasars with S/N $<$ 10 per pixel because we cannot
extract weak emission line features in low S/N spectra.
The redshift range of the 50 spectra is from 0.14 to 1.04 with a mean 
of 0.58.  The average S/N is 19.5 per 0.5 \AA\ binned pixel.


\section{PCA \& Principal Component Spectrum  \NOTE{section b} } \label{sec:b}

\subsection{The PCA Formulation}
\outline{draft}{PCA formulation}
We express a quasar spectrum in Dirac's ``bra ket'' form, $|q_{i}\rangle$,
which is commonly used in Quantum Mechanics and simplifies our
description.
We claim that any quasar spectrum, $|q_{i}\rangle$ ,
is well represented by a reconstructed spectrum, $|r_{im}\rangle$, which
is a sum of the mean and the weighted $m$ principal component spectra:

\begin{equation}
|q_{i}\rangle \sim |r_{i,m}\rangle 
= |\mu\rangle + \sum_{j=1}^{m} c_{ij} | \xi_{j} \rangle
\end{equation}
where $i$ refers to a quasar, $|\mu\rangle$ is the mean quasar spectrum, 
$|\xi_{j}\rangle$ is the $j$th principal component spectrum (PCS), 
and $c_{ij}$ is its weight.
Unlike Quantum Mechanics, the weight, $c_{ij}$, is not a complex number
but a real number.
We found the covariance and correlation matrix of the 50 quasar
spectra in Paper~I and by diagonalizing the covariance matrix, we can
obtain the PCS.
In practice, we found the PCS via a singular value decomposition (SVD)
technique from the 50 quasar spectra. 
The recipes can be found in \citet{francis92} and Paper~I.
Since we designed PCS to be orthonormal:

\begin{equation}
\langle \xi_{i} | \xi_{j} \rangle = \delta_{ij},
\end{equation}
where we define the inner product as:
\begin{equation}
\langle x | y \rangle
=\int_{1020 {\rm \AA } }^{1600 {\rm \AA } }
x(\lambda)\; y(\lambda)\; d\lambda,
\end{equation}
and we can obtain the weights $c_{ij}$ as follows:
\begin{equation}
c_{ij} = \langle q_{i} - \mu | \xi_{j} \rangle .
\end{equation}

\begin{figure}
\includegraphics[angle=270,scale=0.75]{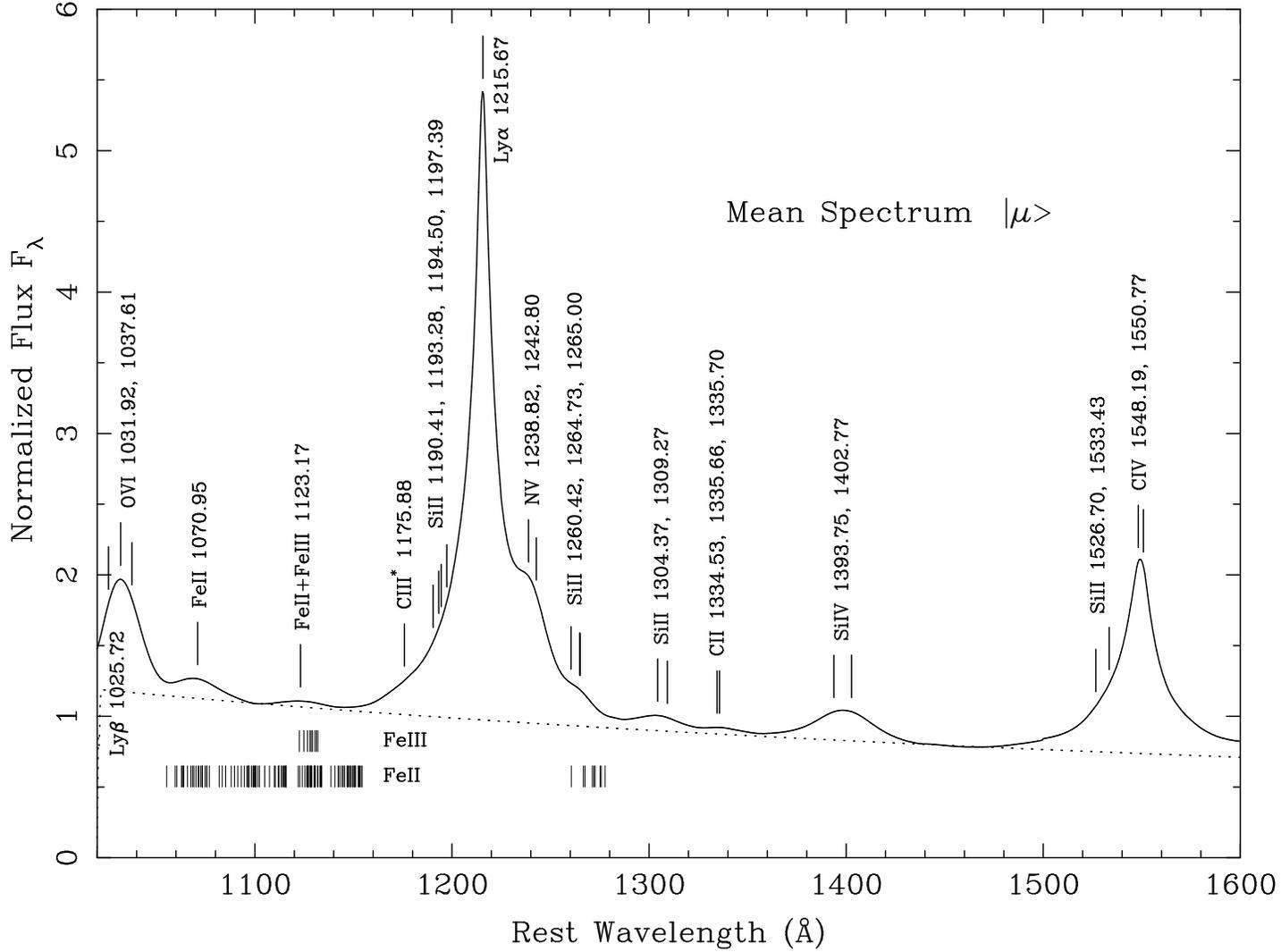}
\caption{
\NOTE{fig:p3}
The mean spectrum of 50 {\it HST} quasar spectra: 
The spectrum is normalized near 1280 \AA.
The wavelengths are taken from \citet{morton91a} except
Fe~II, Fe~III and CIII$^{*}$ lines that are observed wavelenghs 
from \citet{tytler04a}.
The tick marks shown below the spectrum are the wavelengths of 
the Fe~II and Fe~III multiplet.
The dotted line is the power-law continuum apporximation.
Note the emission lines do exit in the Ly$\alpha$ wavelength region,
We also note the wavelength separation of Si~IV doublet at
$\lambda$1400 is relatively large, and makes the line profile
broad.
}
\label{fig:p3}
\end{figure}

\begin{figure}
\includegraphics[angle=0,scale=0.75]{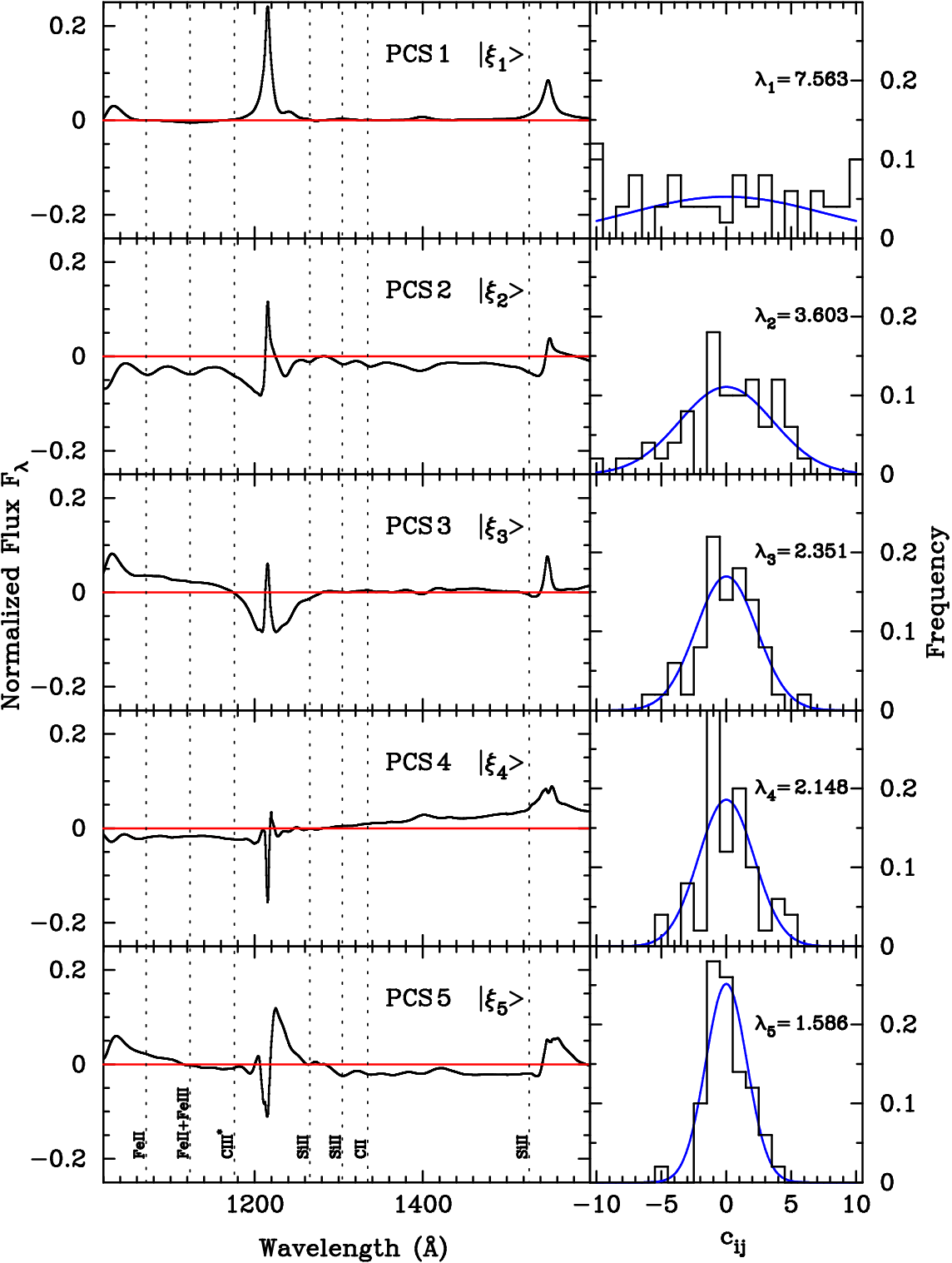}
\caption{
\NOTE{fig:p1}
The first five PCS 1-5 are shown in the left panels with the 
distribution of PCS coefficients $c_{ij}$ on the right.
The first PCS takes the Ly$\alpha$, Ly$\beta$ and the high ionization 
line features (O~VI, Si~IV, C~IV) while the second PCS account for
the low ionization lines (Fe~II, Fe~III, Si~II, C~II).
We expect to have the low ionization lines in the Ly$\alpha$ forest
when a quasar has the negative second PCS coefficient.
We note that PCS are designed to be orthonormal.  
The quasar spectra are normalized near 1280 \AA\ and every
PCS has zero value near 1280 \AA .
}
\label{fig:p1}
\end{figure}

\outline{draft}{Principal Component Spectra, figure}
We show the mean spectrum in Figure \ref{fig:p3} and
the first 10 PCS and the distribution of their weights
in Figure \ref{fig:p1} and \ref{fig:p2}.
Although wavelength coverage, normalization, resolution,
and the number of quasars are different, 
the general trend of PCS is similar to that of \citet{francis92} 
with the exception of the third PCS.
The reason for the exception is that 
the third PCS of \citet{francis92} takes into account BAL features, 
while we don't have such features since we removed BAL quasars.  
Our third PCS has sharp emission line features for Ly$\beta$/O~VI,
the core of Ly$\alpha$, and C~IV emission lines, but broad negative
features around the Ly$\alpha$ emission line.
The continua of the third, fourth and fifth PCS have a slope,
and the fifth PCS includes the asymmetric feature for the strong 
emission lines.
The fifth PCS is similar to the second PCS in that P-Cygni profiles are
seen but they are very broad, and there are no low ionization emission 
line features blueward of the Ly$\alpha$ emission line.
The sixth and seventh PCS carry some information on low ionization 
weak emission lines, but the spectrum features are getting noisy.
The eighth and higher order PCS have high frequency wiggles which 
no longer correspond to any physical emission lines and
are probably due to fitting errors.
We used continua fitted spectra which are free from photon noise,
but still they are likely to suffer fitting errors of 
at least a few percent
as we can see in Figures \ref{fig:p7}-\ref{fig:p11}.
The mean spectrum and the first 10 PCS are available on-line from
the Paper~I.
Next, let us discuss the contribution from each PCS quantitatively.

\subsection{Quantitative Assessment of PCA Reconstruction}
\outline{draft}{Variance, table}
We make use of the residual variance to assess the goodness of PCA 
reconstruction.
The residual variance of a reconstructed quasar spectrum with $m$ components 
is a sum of the squares of the difference between the quasar 
spectrum, $|q_{i,}\rangle$, and the reconstructed spectrum, $|r_{i,m}\rangle$:

\begin{equation}
\langle q_{i} - r_{i,m} | q_{i} - r_{i,m} \rangle 
= \sum_{i=m+1}^{N} c_{ij}^{2},
\end{equation}
where $N$ is the total number of independent PCS; either
the number of quasar spectra or the total number of pixels.
In our case, $N$ is the number of quasar spectra (50) since
it is smaller than the total number of pixels (1167).
For the overall contribution from $m$th PCS, we define the 
residual variance fraction, f($j$), as follows:

\begin{equation}
f(j)= \frac{\displaystyle \sum_{i=1}^{N} c_{ij}^{2}}
     {\displaystyle \sum_{i,j=1}^{N} c_{ij}^{2}}.
\end{equation}

We redefine the square root of the eigenvalue of the 
$j$th PCS as follows:

\begin{equation}
\lambda_{j}=\frac{1}{N-1}\sqrt{\sum_{i=1}^{N}c_{ij}^{2}}.
\end{equation}

\begin{figure}
\includegraphics[angle=0,scale=0.75]{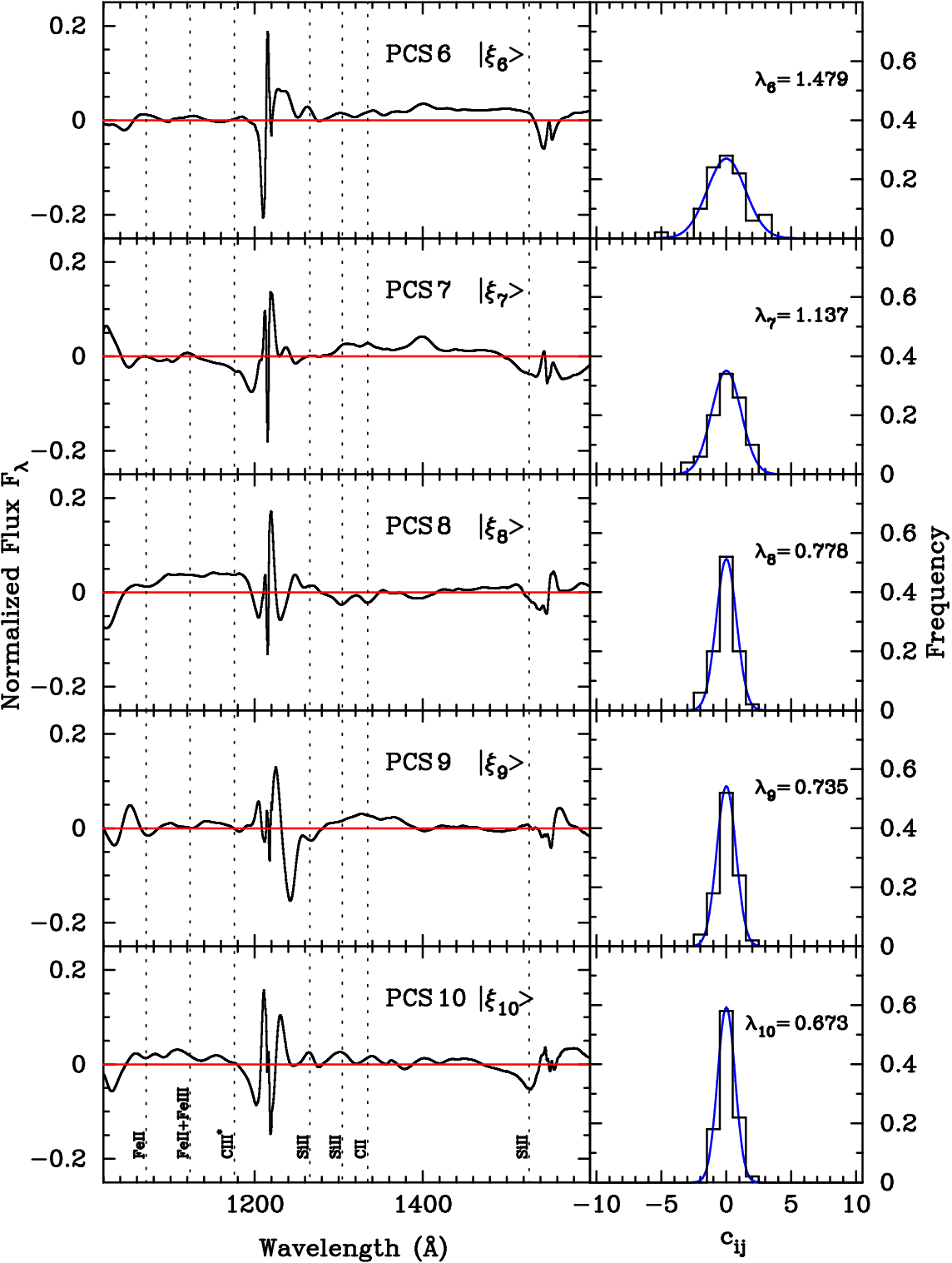}
\caption{
\NOTE{fig:p2}
The same as Figure \ref{fig:p1} but the second five PCS 6-10.
From 8th and higher PCS, the features in PCS become noisy and 
no longer have physical correspondence.  
They are probably fitting errors.
}
\label{fig:p2}
\end{figure}

\outline{draft}{$\lambda$ advantage }
Then we wish to use $\lambda_{j}$ to describe the probability 
distribution function (PDF) of the PCS coefficients.
The PDF of the weights is not necessarily a Gaussian, but as shown
in Figures \ref{fig:p1} and \ref{fig:p2}, the PDF can be well 
represented by a Gaussian.  We also note that by design: 

\begin{equation}
\sum_{i=1}^{N}c_{ij}=0
\end{equation}
for any $j$.
Thus, the PDF of weights is characterized by just one 
parameter, $\lambda_{j}$. 
The probability of having a weight, 
$c_{ij}$, in an interval, $-x_{0} \leq c_{ij} \leq x_{0}$ can be 
expressed as:

\begin{equation}
\displaystyle
P(-x_{0} \leq c_{ij} \leq x_{0}) = 
\int_{-x_{0}}^{x_{0}}\frac{1}{\sqrt{2\pi}\lambda{j}} 
e^{-\frac{x^{2}}{2\lambda_{j}^{2}}}dx.
\end{equation}

Naturally, it would be convenient if we standardize the weight, $c_{ij}$,
by $\lambda_{j}$. 
We can then rewrite a quasar spectrum as:
\begin{equation}
|q_{i}\rangle \sim |r_{i,m}\rangle 
= |\mu\rangle + \sum_{j=1}^{m} \lambda_{j} \sigma_{ij} | \xi_{j} \rangle 
\label{eqn:pcs}
\end{equation}
where the $\sigma_{ij}$ are the standardized PCS coefficients which 
represent the deviation from the mean spectrum of the $j$th PCS of 
quasar $i$.
The PDF of the $\sigma_{ij}$ is a normal distribution, so we can immediately
tell how far and how different the quasar spectrum would be from the 
mean spectrum.
This standardization of the weights simplifies the discussion of
the variety of quasar spectra in \S \ref{sec:i}.

\outline{draft}{$\lambda$ advantage 3}
Another advantage of using $\lambda_{j}$ is that we can simplify the 
residual variance fraction f(j): 

\begin{equation}
f(j)=\frac{\lambda_{j}^{2}}{\displaystyle \sum_{j=1}^{N}\lambda_{j}^{2}}. 
\end{equation}


\outline{draft}{Goodness Assessment}
The values of f($j$) are listed in Table \ref{tbl:a}. 
The first, second and third PCS account for 63.4\%, 14.5\% and 6.2\% 
of the variance respectively.
In total, the first three PCS account for 84.3\% of the variance of 
the 50 quasars in our sample. 
These fractions depend on the normalization and the wavelength
coverage.
In the literature, \citet{francis92} report that the first 
three PCS account for 75\% ($\lambda\lambda 1150-2000$), and
\citet{shang03a} show that the first three PCS carry 80\% 
($\lambda\lambda1171-2100$) of the variance.
Both groups normalized flux by the mean flux, while this paper
normalizes by a flux value near 1280~\AA.
Our value of 84.3\% is slightly higher than the above numbers probably 
because we removed the BAL quasars, which are certainly a source of 
variance.
In addition, we used fitted smooth continua to the Ly$\alpha$ forest, 
while they used the observed raw Ly$\alpha$ forest
which obviously contains a large variance \citep{tytler04c}.
As shown in Table \ref{tbl:a}, the contribution from each 
PCS component to the variance rapidly decreases with order $m$.
It becomes less than 1\% after the 8th PCS and then 
stays the same. 
With seven PCS components, 96\% of the variance has already been 
accounted for. 
As is seen in the PCS features in Figure \ref{fig:p2}, 
the remaining 4\% of the variance is probably due to fitting errors.
\section{Artificial Spectra  \NOTE{section f} } \label{sec:f}
\outline{draft}{Active use of PCS}
We can use PCS 
to generate artificial spectra.
Artificial spectra can be useful in testing the detection of quasars and DLAs,
in flux calibration, in continuum fitting, and in cosmological simulations.
By assigning PCS coefficients randomly from known PDFs,
we can generate artificial quasar spectra.
As we have discussed in \S \ref{sec:b}, the PDF of the $j$th PCS
coefficient is well represented by a Gaussian with a mean of 0, and 
a standard deviation of $\lambda_{j}$.
If we then sum up the PCS with these coefficients (equation \ref{eqn:pcs}), 
we can create a set of artificial quasar spectra.

\outline{draft}{Use of Artificial Quasar Spectra}
Noiseless quasar spectra are of great use in IGM studies, since
at high redshift, it is difficult to see the continuum in the Ly$\alpha$
forest.
Even at redshift 2, pixels in the Ly$\alpha$ forest hardly reach
the continuum level with FWHM=250km/s \citep{tytler04a}.
Artificial quasar spectra can thus be useful to predict the shape of
continuum in the Ly$\alpha$ forest (Paper~I) and to calibrate
continuum fitting to the Ly$\alpha$ forest \citep{tytler04c,tytler04a}.
They would be also useful to test the detection limit of the high redshift
quasar survey since the Ly$\alpha$ emission can possibly boost the 
brightness by 0.15 magnitude.
We have generated 10,000 artificial quasar spectra using the
first seven PCS for a more realistic representation of the quasar
spectra.
We concluded that the features seen in PCS greater than eighth are noise,
and we did not include higher order PCS.
We will provide artificially generated spectra to the community upon 
request.

\section{PCA Classification  \NOTE{section d} } \label{sec:i}

\subsection{Introduction of Five Classes}
\begin{figure}
\includegraphics[angle=0,scale=0.75]{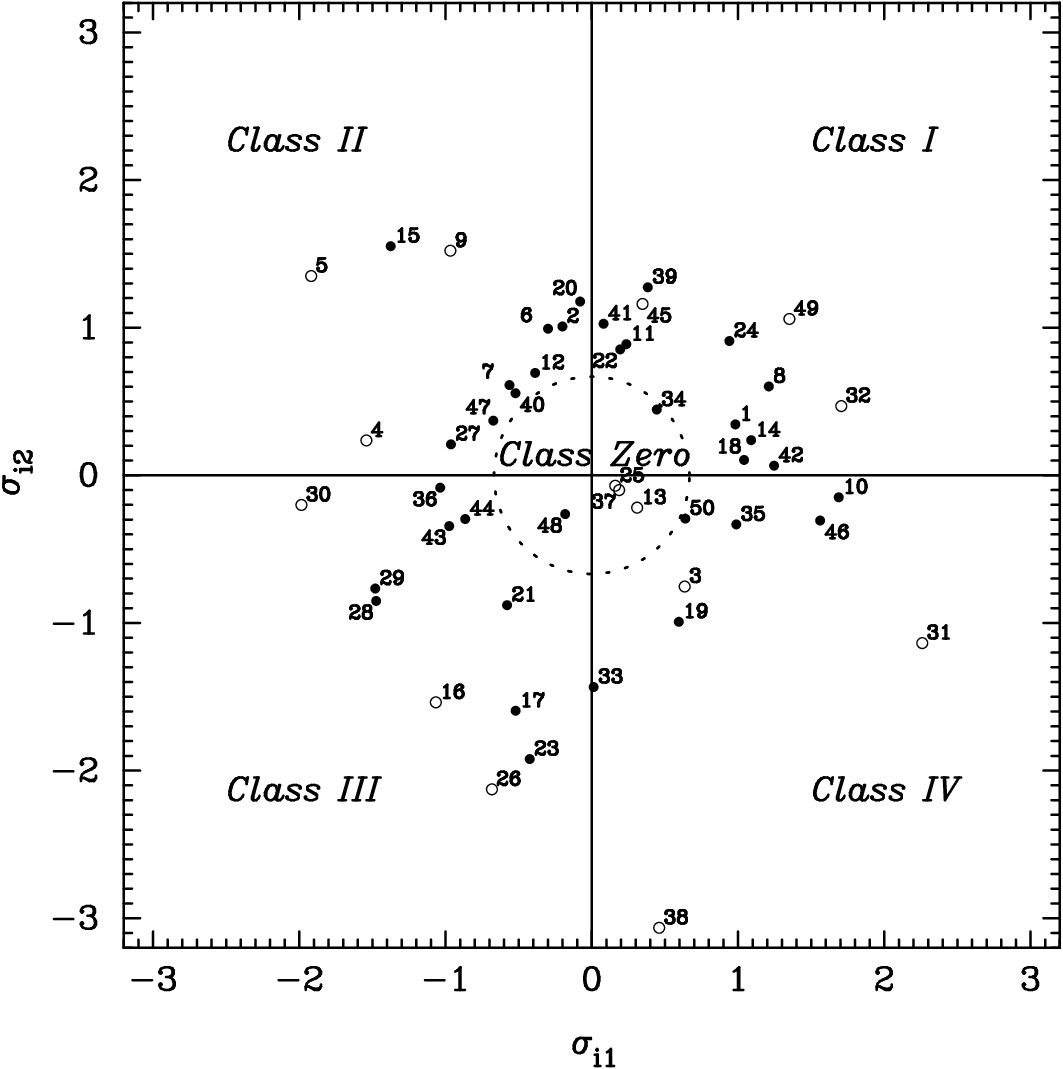}
\caption{
\NOTE{fig:p5}
The distribution of standardized first two PCS coefficients 
for 50 quasars.
The small number besides each point is quasar number $i$ which 
is listed in Table \ref{tbl:e}. 
For quasars with open circles, we have quasar spectra shown 
in Figure \ref{fig:p7}-\ref{fig:p11}.  
We intentionally chose extreme cases to show a wide variety
of quasar spectra.
The the dotted circle has the radius of 0.668.
We divide this plane into the five zones and define the five
classes.  
We define {\it Class Zero} as the quasars which have radius less 
than or equal to 0.668 and {\it Class I-IV} as the ones which 
have radius greater than 0.668 and are in quadrants {\it I-IV}
respectively.
}
\label{fig:p5}
\end{figure}

\outline{draft}{Parameters for Classification}
This paper attempts to classify quasar spectra quantitatively using
our standardized PCS coefficients, $\sigma_{ij}$.
We note that there is no discrete classification of quasar spectra
and that they vary continuously.
However, this classification will help us in fitting the continua to 
the Ly$\alpha$ forest spectrum.
We use the first two PCS coefficients to demonstrate the variety of
emission lines and continua.
As we have seen in \S \ref{sec:b} and Table \ref{tbl:a}, the first two PCS
coefficients account for 77.9\% of the variance and represent
the overall shape of the quasar spectrum.
We introduce polar coordinates as follows:

\begin{eqnarray}
         r_{i}& = & \sqrt{\sigma_{i1}^{2} + \sigma_{i2}^{2}}\\
\tan\theta_{i}& = & \frac{\sigma_{i2}}{\sigma_{i1}}
\end{eqnarray}
where $r_{i}$ represents the deviation from the mean spectrum, and the 
angle $\theta_{i}$ tells us about the profiles of the emission lines. 
We divide the $\sigma_{i1}$ vs. $\sigma_{i2}$ diagram into five zones
and introduce five classes of spectral types that enable us to discuss
the shape of continua qualitatively.
The main goal of this classification is to differentiate the families of 
quasar spectra.
First, we define {\it Class Zero} for those close to the
mean spectrum in shape.
We define {\it Class I-IV} corresponding to the quadrant {\it I-IV} 
in the $\sigma_{i1}$ vs. $\sigma_{i2}$ diagram, shown 
in Figure \ref{fig:p5}.
The probability of $r \leq r_{0}$ is:

\begin{eqnarray}
P(r \leq r_{0})& = & \int_{0}^{r_{0}}r e^{-\frac{r^{2}}{2}} dr\\
               & = &1 - e^{-\frac{r_{0}^{2}}{2}}.
\end{eqnarray}

Now, we wish to design the fraction of the five classes to be equal, 
namely 0.2 each.
We find $r_{0}=0.668$ gives $P(r \leq r_{0})=0.2$,
so we define {\it Class Zero} for a quasar spectrum that has
$r \leq $ 0.668. 
For quasar spectra with $r > 0.668$,  we define the 
classes {\it I} through {\it IV} corresponding to the quadrants 
first through fourth on the $\sigma_{i1}$ vs. $\sigma_{i2}$ 
diagram (Figure \ref{fig:p5}).

Our standardized PCS coefficients are plotted on the
$\sigma_{i1}$ vs. $\sigma_{i2}$ diagram in Figure \ref{fig:p5}
where the dotted circle is the circle with radius $r_{0}=0.668$. 
The small numbers noted beside the points are the quasar identification
numbers $i$ that are listed in Table \ref{tbl:e}. 

\subsection{Demonstration of Five Classes}
\begin{figure}
\includegraphics[angle=270,scale=0.75]{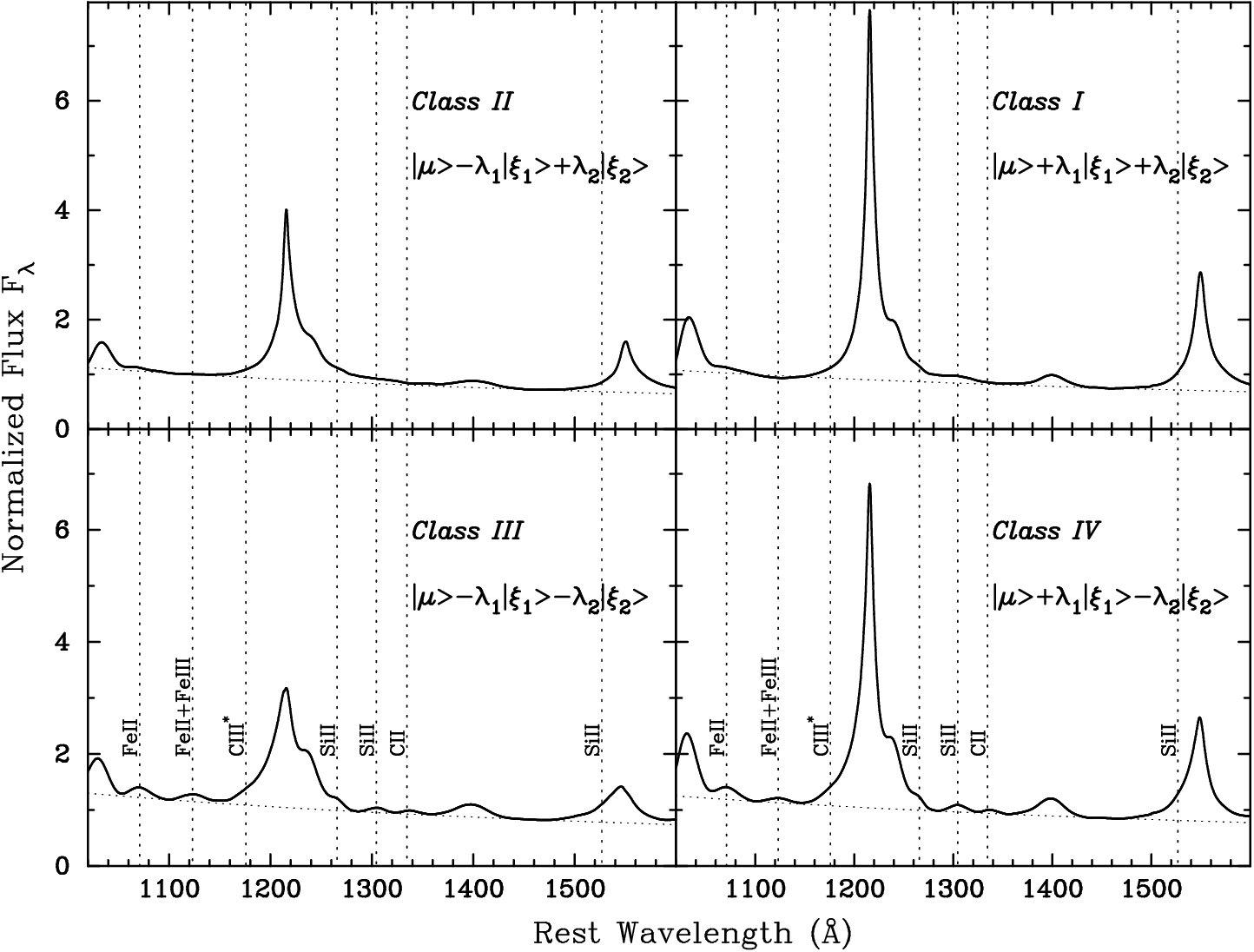}
\caption{
\NOTE{fig:p4, notations will be given in the figure}
The illustration of four classes.
We artificially generated these four spectra using the first two PCS.
We chose $\pm \sigma_{i1}$ and $\pm \sigma_{i2}$  to generate
the four spectra, and equation is given in the legend where
$| \mu \rangle$ is the mean spectrum, $\lambda_{1}$ and $\lambda_{2}$ are
the square root of the first two eigenvalues, and $| \xi_{1} \rangle$ 
and $| \xi_{2} \rangle$ are the first two PCS.
The wavelengths of low ionization lines are shown as vertical
dotted lines.
Note that the four spectra are plotted in the same scale, and 
the emission line peak contrast with the continuum characterizes 
the classification.
{\it Class I} and {\it II} do not have any weak emission lines
and the power-law continuum fit (smooth dotted line) is a good
approximation of the continuum in the Ly$\alpha$ forest while
{\it Class III} and {\it IV} show prominent low emission lines.
}
\label{fig:p4}
\end{figure}

\outline{draft}{Examples}
We show the mean spectrum in Figure \ref{fig:p3}.  By definition
the mean spectrum $| \mu \rangle$ has r=0, and naturally it belongs 
to {\it Class Zero}.  
In Figure \ref{fig:p4} we show the artificially generated 
four classes, {\it I-IV}, of the quasar spectra to illustrate their 
typical spectral shape.
They are the sum of the mean and the first
two PCS with $\sigma_{i1}=\pm1$ and $\sigma_{i2}=\pm1$.
The generated four spectra of {\it Class I-IV} have 
angles $\theta$:  $45^{\circ}, 135^{\circ}, 225^{\circ}$ 
and $315^{\circ}$ respectively, and they all have $r=\sqrt{2}$.
The four spectra are plotted on the same scale in Figure \ref{fig:p4} 
so that we can see the contrast of the emission lines with the
continuum in a uniform manner.

\outline{draft}{Examples of Observed Spectra}
In Figure \ref{fig:p7}-\ref{fig:p11}, we show three observed spectra from 
each class where we intentionally chose the extreme cases for 
{\it Class I-IV}.  
Quasars are plotted at rest frame wavelengths with
the luminosities that are calculated by using the cosmological parameters
from the first year WMAP observation 
\citep[h=0.71, $\Omega_{m}$=0.27, $\Omega_{\Lambda}$=0.73 ;][]{spergel03a},
and a flat universe is assumed.
The smoothed line on the spectrum is the fitted continuum, and the solid
straight line is the power-law continuum fit.
The vertical dotted lines show the wavelengths of emission lines in the
Ly$\alpha$ forest and the low ionization lines redward of the Ly$\alpha$
emission line.
The quasar numbering in Figure \ref{fig:p7}-\ref{fig:p11}
is the same as in Figure \ref{fig:p5} so that we can visualize where
the quasar spectrum is in the $\sigma_{i1}$ vs. $\sigma_{i2}$ diagram.
In Table \ref{tbl:e}, quasars are sorted by classes, and the equivalent 
width of the emission lines are listed.


\subsection{The Characteristics of the Five Classes}
\outline{draft}{First PCS and the second PCS}
The characteristics of the first two PCS directly reflect on
the five classifications, thus
let us take a close look at the first two PCS in Figure \ref{fig:p1}.
The first PCS carries the sharp and strong lines --
Ly$\alpha$, Ly$\beta$ and high ionization emission line
features (O~VI, N~V, Si~IV, C~IV).
The second PCS has low ionization emission line features:
Fe~II and Fe~III in blueward of the Ly$\alpha$ emission line, 
Si~II and C~II redward. 
Their profiles are broad and rounded.
In the second PCS, the flux values of low ionization emission lines
and the strong Ly$\alpha$, C~IV emission lines have opposite sign,
meaning they are anticorrelated.
In addition to that, Ly$\alpha$ and C~IV emission lines have P-Cygni
profiles which introduces asymmetric profiles to these emission lines.

\outline{draft}{Class III \& IV : Second PCS coefficient}
By definition, the first two PCS engage the correlation 
between emission lines. 
Since we are particularly interested in the
profiles of emission lines in the Ly$\alpha$ forest, let us
look at low ionization lines first.
If a quasar shows prominent low ionization lines 
redward of the Ly$\alpha$ emission line 
(Si~II $\lambda\lambda 1260,1304$, C~II $\lambda1334$),
it should have a negative second PCS coefficient, and
we should expect to have prominent Fe~II $\lambda1070$ and
Fe~III $\lambda1123$ in the Ly$\alpha$ forest.
Thus such a quasar should belong to either {\it Class III} or {\it IV}.
As a consequence, these two classes have the largest equivalent 
widths of these low ionization lines among the five classes
as seen in Table \ref{tbl:e}.

\outline{draft}{Class I \& II : First PCS coefficient}
If another quasar has sharp and strong Ly$\alpha$, Ly$\beta$ 
and high ionization lines (N~V, Si~IV, C~IV), it should have a 
positive first PCS coefficient and belong to {\it Class I} or 
{\it IV}.
The normalized flux of the Ly$\alpha$ emission peak
and the ratio of the Ly$\alpha$ and N~V peak flux are the highest 
for {\it Class I} among the five classes.
\outline{draft}{Class I-IV : PCS coefficient}
We can differentiate {\it Class I} and {\it II}, or 
{\it Class III} and {\it IV} by combining the above characteristics.
As we expect, the diagonal classes have the opposite 
characteristics. 
For example, {\it Class I} has sharp and high ionization lines while
{\it Class III} has broad and rounded low emission lines.

\outline{draft}{Key Point}
In practice, the key point of finding the continuum level in the Ly$\alpha$ 
forest is to seek the low ionization lines 
(Si~II $\lambda\lambda 1260,1304$, C~II $\lambda1334$) 
and their profiles redward of the Ly$\alpha$ emission.
If we see them, we should expect to have similar profiles of Fe~II
$\lambda 1070$ and Fe~II $\lambda1123$ lines in the Ly$\alpha$ forest.
If we do not see them, we can expect the continuum to be flat in the
Ly$\alpha$ forest, and the power-law extrapolation from the redward
of the Ly$\alpha$ emission to be a good approximation.
We will discuss the accuracy of the power-law extrapolation in the
next section.


\section{Mean Flux $\langle F \rangle$ and 
Flux Decrement D$_{A}$ \NOTE{section g} } \label{sec:g}
In performing a precise measurement of the IGM flux decrement D$_{A}$,
it is essential to include the effect of weak emission lines found
in the Appendix A. 
This contribution is not negligible, and will be quantitatively
assessed.

\subsection{A Brief History of the Flux Decrement D$_{A}$}
\outline{draft}{A Brief History of Da}
\citet{gunn65a} predicted a flux decrement due to foreground 
neutral hydrogen in the IGM, namely the Ly$\alpha$ forest.
\citet{oke82} first defined and measured the flux decrement, D$_{A}$.
\citet{schneider91} introduced the Ly$\alpha$ forest wavelength 
interval for D$_{A}$ as $\lambda\lambda1050-1170$, and it has 
been widely used since \citep{zuo93a,kennefick95a,spinrad98a}.
\citet{madau95} and \citet{mcdonald01b} have used D$_{A}$ to estimate 
the UV background. 
The flux decrement of the high redshift IGM probes the 
reionization epoch of the universe \citep{loeb01}.
The current estimate of the reionization epoch from the IGM 
is around $z \sim 6-7$ \citep{becker01,djorgovski01,fan03a}, while 
the first year WMAP satellite estimates $z \sim 20$ 
\citep{spergel03a}. The discrepancy is yet to be resolved or 
explained \citep{cen03a}.

\outline{draft}{Da in Cosmology}
A precise measurement of the flux decrement, D$_{A}$, is of 
great importance for studies of the IGM
\citep{rauch98} because it is very sensitive to the cosmological 
parameters, $\sigma_{8}$ (the amplitude of 
the mass power spectrum) and $\Omega_{\Lambda}$, as well as to the 
UV background intensity \citep{tytler04a,jena04a}.
However, it is this sensitivity that makes the D$_{A}$ measurement
a major source of error \citep{hui99a,croft02b}.


\subsection{The Mean Flux Correction Factor $\delta F$}

\outline{draft}{Da Definition}
The flux decrement is defined as:

\begin{equation}
D_{A}=1 - \int_{ 1050 {\rm \AA} }^{ 1170 {\rm \AA}} 
\frac{f_{\lambda}(\lambda; Observed)}
{f_{\lambda}(\lambda; Continuum)}d\lambda
\end{equation}

Thus, what we are measuring is the mean flux:

\begin{equation}
\langle F \rangle=\int_{ 1050 {\rm \AA} }^{1170 {\rm \AA}} 
\frac{f_{\lambda}(\lambda; Observed)}
{f_{\lambda}(\lambda; Continuum)}d\lambda
\end{equation}

\begin{figure}
\includegraphics[angle=0,scale=0.8]{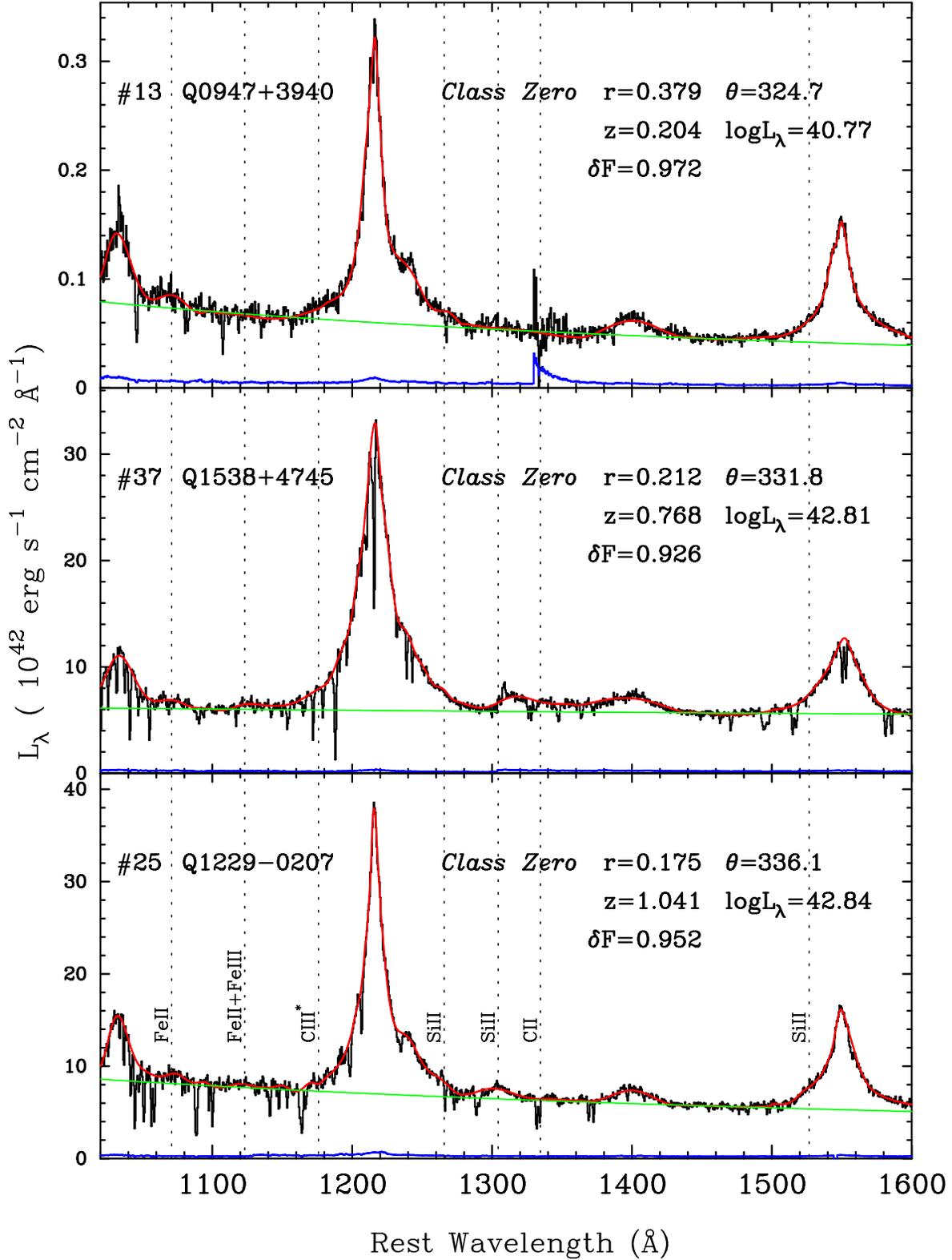}
\caption{
\NOTE{fig:p7}
{\it Class Zero} ( r $\leq$ 0.668) :
The smooth line on the spectrum is the fitted continuum, and 
the straight solid line is the power-law continuum fit.
The line at the bottom is 1-$\sigma$ error of the spectrum.
The vertical dotted lines show the wavelengths of emission
lines in the Ly$\alpha$ forest and low ionization lines
redward of the Ly$\alpha$ emission line.
The luminosity, L$_\lambda$, is measured at $\lambda$1280 where
we normalized the spectra.
The emission lines in the Ly$\alpha$ forest are barely seen
in the three spectra.
A discontinuity of the spectrum is seen in the middle of the spectrum
in Q0947+3940 and Q1538+4745.  
These are due to the different gratings used in the observations
to cover the wide range of wavelength. 
We joined them together by taking the weighted mean, but it does
not always give a smooth solution.
This discontinuity could be another source of fitting errors.
}
\label{fig:p7}
\end{figure}

\begin{figure}
\includegraphics[angle=0,scale=0.8]{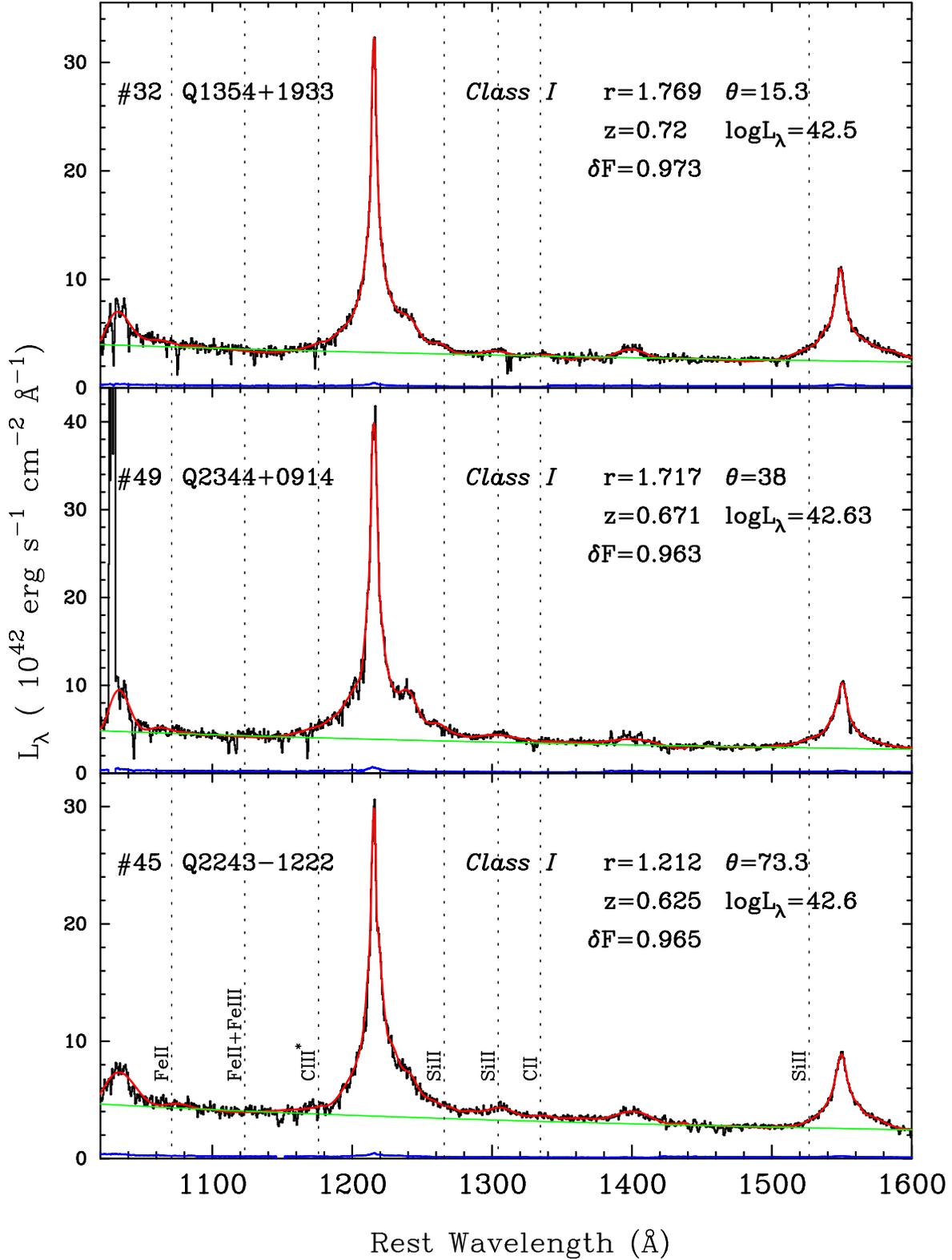}
\caption{
\NOTE{fig:p8}
{\it Class I} ( r $>$ 0.668, $ 0 \leq \theta < 90^{\circ} $):
The Ly$\alpha$, Ly$\beta$ and high ionization emission lines are all
very sharp and high contrast with the continuum.
The low ionization lines, Si~II at $\lambda\lambda1260,1304$, are 
barely seen.
We see no clear sign of emission lines in the Ly$\alpha$ forest, and
the power-law continuum is a good approximation. 
The contrast of Ly$\alpha$ emission peak with the continuum is the
highest among the five classes.
In Q2344+0914, there is a huge spike at 1030 \AA .
It is due to an emission line caused by the Earth's atmosphere.
We removed these spikes when we fitted the continuum.
}
\label{fig:p8}
\end{figure}

\begin{figure}
\includegraphics[angle=0,scale=0.8]{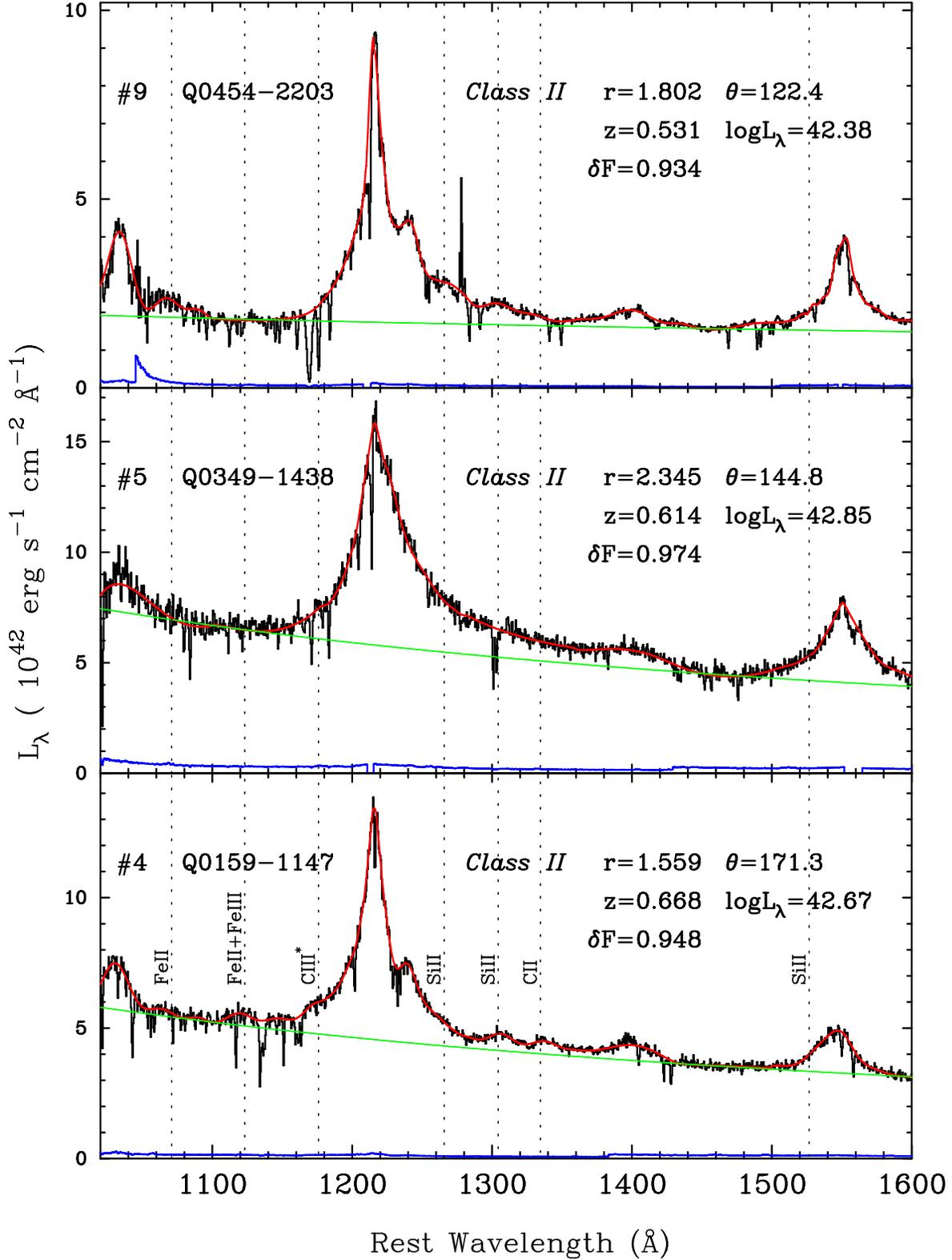}
\caption{
\NOTE{fig:p9}
{\it Class II} ( $ r > 0.668$, $ 90^{\circ} \leq \theta < 180^{\circ} $):
The Ly$\alpha$ emission peak has a moderate contrast, 4-6, 
with the continuum, and the emission peak ratio of Ly$\alpha$ and N~V is 2-3.
The Ly$\alpha$ emission has long tails, and the blueward tail of the 
fitted continuum does not meet with the power-law continuum until 1120 \AA ,
which is 50 \AA\ below the D$_{A}$ wavelength definition, 1170 \AA.
In an extreme case, Q0349-1438 ($r=2.345$), the
tail of Ly$\alpha$ emission wings last about 100 \AA .
Ly$\beta$/O~VI, Ly$\alpha$, and C~IV emission profiles are
triangular, and Si~IV $\lambda1393$ is very broad and rounded.  
There is no clear separation between Ly$\alpha$ and N~V emission lines
and no sign of low ionization emission lines.
}
\label{fig:p9}
\end{figure}

\begin{figure}
\includegraphics[angle=0,scale=0.8]{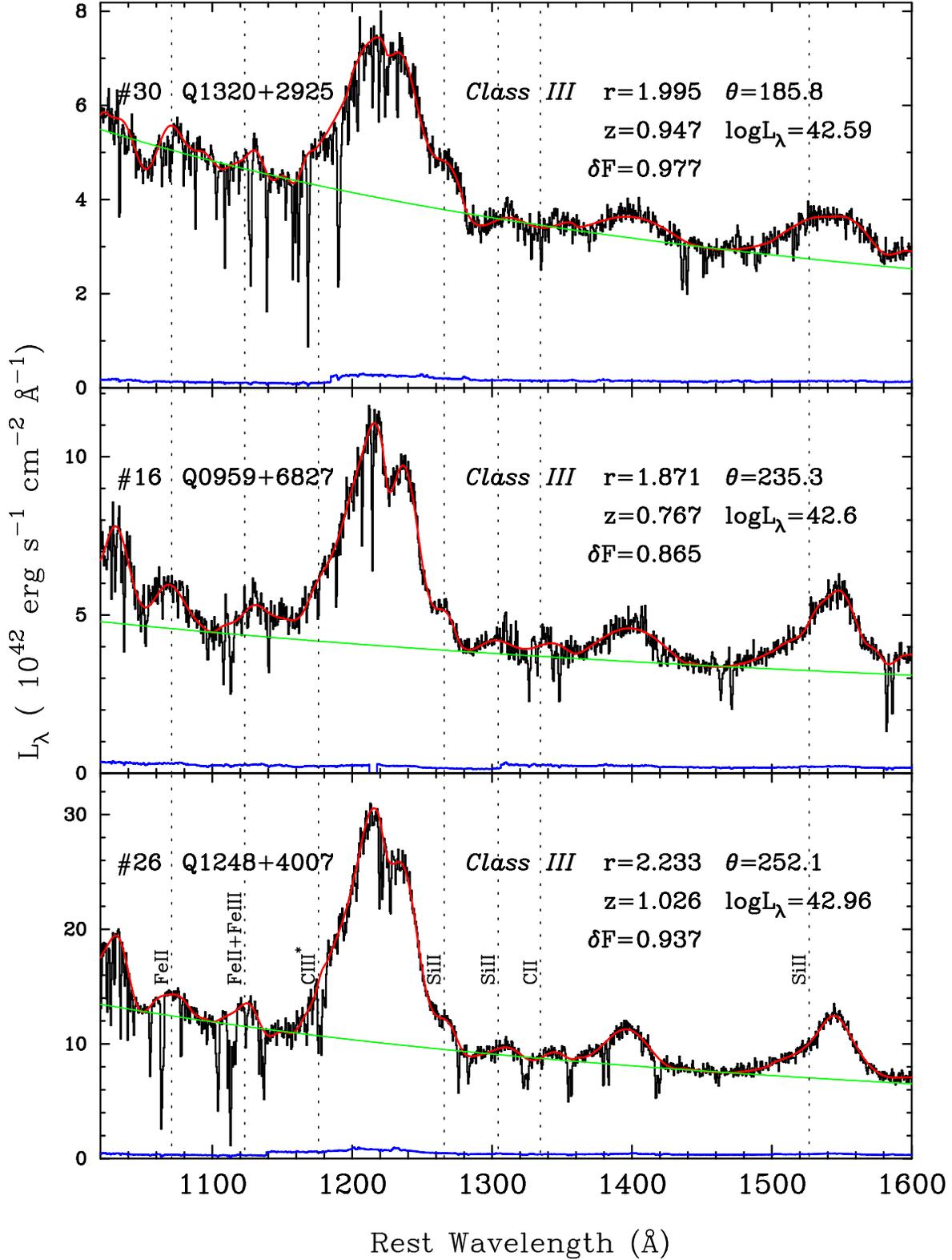}
\caption{
\NOTE{fig:p10}
{\it Class III} ($ r > 0.668$,  $ 180^{\circ} 
\leq \theta < 270^{\circ} $ ):
The emission line profiles are all broad and rounded.
Fe~II and Fe~III lines are clearly seen in the Ly$\alpha$ forest.
The contrast of the Ly$\alpha$ emission line peak with the continuum
is the lowest, 2-4, among the five classes.
The ratio of Ly$\alpha$ emission peak to N~V is also the lowest: 1-2.
The C~IV profile is asymmetric probably because of the contribution
from the Si~II $\lambda1526$ hidden inside the blueward tail of the
C~IV line.
}
\label{fig:p10}
\end{figure}

\begin{figure}
\includegraphics[angle=0,scale=0.8]{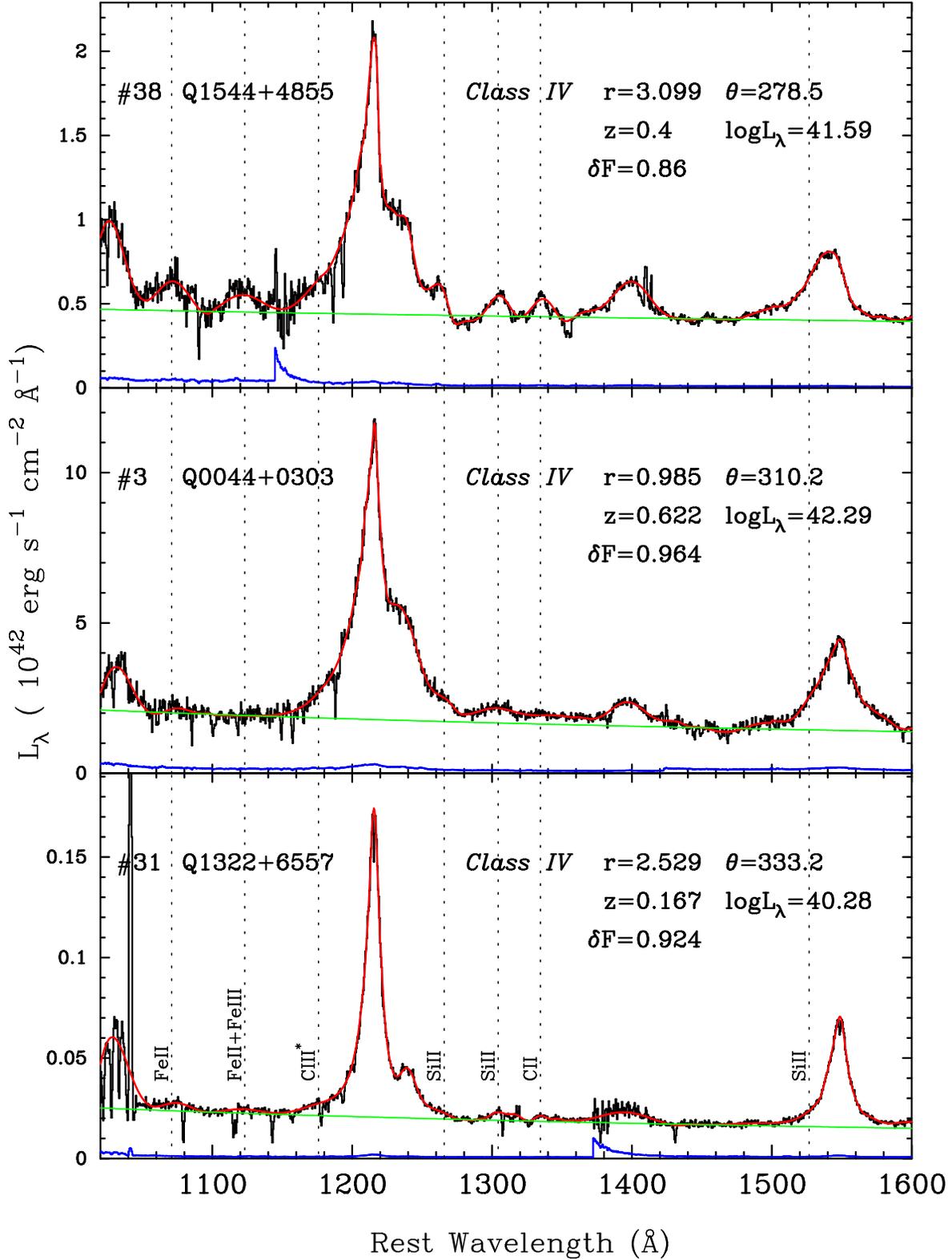}
\caption{
\NOTE{fig:p11}
{\it Class IV} ($ r > 0.668$,  $ 270^{\circ} \leq \theta < 360^{\circ}$):
The high contrast with the continuum and sharp emission profile are the
characteristics of this class.
Q1544+4855 is an extreme case that has r=3.099.
Since it has $\theta=278.5$, it is very close to {\it Class III},
but it has the characteristics of {\it Class IV}: high Ly$\alpha$ 
emission line peak contrast and the very sharp line profiles.
Unlike a {\it Class III} quasar, the peaks of low ionization line profile 
are very sharp and not rounded.
Q1322+6557 has a spike at $\lambda1040$ that is due to an 
atmospheric emission line.  
The fitted spectrum removed the spike for analysis.
}
\label{fig:p11}
\end{figure}

However, the unabsorbed continuum level is not seen in the Ly$\alpha$
forest and the power-law extrapolation from redward of Ly$\alpha$
emission has been used as a continuum.
In fact, what is reported in the literature as the mean flux is:

\begin{equation}
\langle F_{Power-Law} \rangle=\int_{ 1050 {\rm \AA} }^{1170 {\rm \AA}} 
\frac{f_{\lambda}(\lambda; Observed)}
{f_{\lambda}(\lambda; Power\!-\!Law)}d\lambda
\end{equation}
which is not exactly the same as $\langle F \rangle$ since 
the power-law is a crude approximation of the continuum in the 
Ly$\alpha$ forest as we have seen in \S \ref{sec:i}, Appendix A,
and Figures \ref{fig:p7}-\ref{fig:p11}.
We wish to introduce a correction factor $\delta F$ :
\begin{equation}
\label{eqn:dF}
\delta F = \int_{ 1050 {\rm \AA} }^{1170 {\rm \AA}} 
\frac{f_{\lambda}(\lambda;Power\!-\!Law)}
{f_{\lambda}(\lambda; Continuum)}d\lambda 
\end{equation}
so that we can estimate the true mean flux $\langle F \rangle $ from
the reported mean flux $\langle F_{Power-Law} \rangle$:

\begin{equation}
\displaystyle
\langle F \rangle = \langle F_{Power-Law} \rangle \cdot \delta F .
\end{equation}

\outline{draft}{Slope $\alpha_{\nu}$}
To calculate $\delta F$, we need to find the power-law extrapolation 
from the redward Ly$\alpha$ emission.
Since our wavelength range is limited and not as large as other 
survey data, it is not as easy to extrapolate.
Moreover, we have a series of emission lines, and it is hard
to define a continuum level with no emission lines.
For example, as shown in Figure \ref{fig:p10}, {\it Class III} quasars have 
emission lines throughout this wavelength range.
However, we have a fitted continuum in the Ly$\alpha$ forest, and
we take advantage of it.
We choose two points and find a power-law fit which runs through
these two points.  
We choose one from blueward ($\lambda1100$) and the other 
from redward ($\lambda1450$) of the Ly$\alpha$ emission.
Then, we can have an inter- and extra-polated power-law continuum
and its exponent $\alpha_{\nu}$, for $f_{\nu} \propto \nu^{\alpha_{\nu}}$. 
The average of $\alpha_{\nu}$ is -0.854 with the standard deviation of
0.507.
The power-law continua show that they are all sensible first order
approximations which well represent the continua in the Ly$\alpha$ 
forest. 
The power-law continua are shown in Figure \ref{fig:p7}-\ref{fig:p11}.

\begin{figure}
\includegraphics[angle=270,scale=0.75]{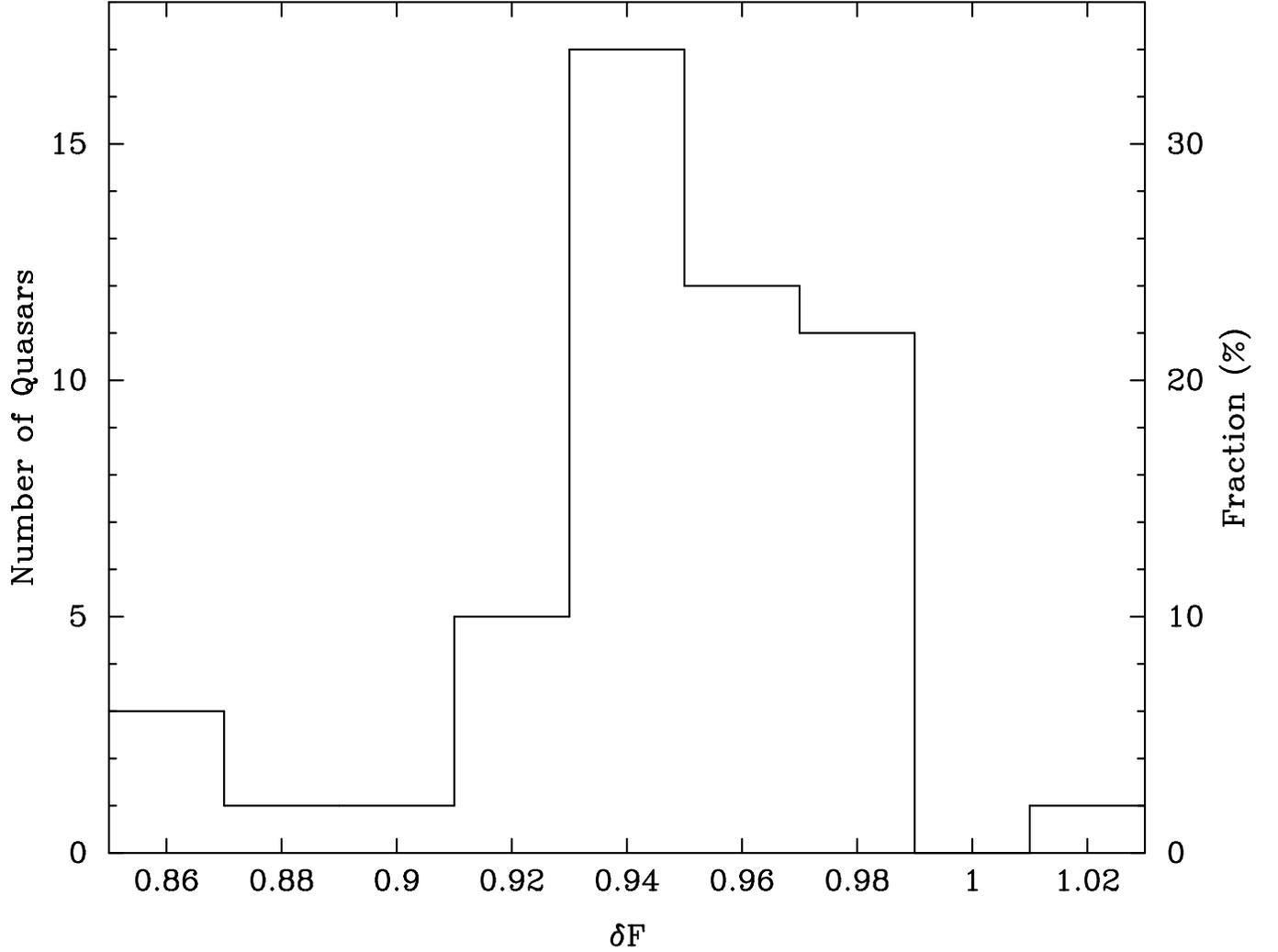}
\caption{
\NOTE{fig:p12}
The distribution of the mean flux correction factor $\delta F$.
The mean is 0.947 with the standard deviation of 0.031.
The distribution is asymmetric with long tail toward small 
$\delta F$ values.
{\it Class III} and {\it IV} quasars have prominent emission 
lines in the Ly$\alpha$ forest, and tend to have small $\delta F$
values.
Two extreme spectra for low $\delta F$ are shown in Figure 
\ref{fig:p10} for Q0959+6927 and Figure \ref{fig:p11} for
Q1544+4855.
}
\label{fig:p12}
\end{figure}

\outline{draft}{Da Correction}
The calculated $\delta F$ is listed in Table \ref{tbl:e} and the 
average of $\delta F$ is 0.947, with a standard deviation of 
0.031. 
This result means that the power-law approximation misses 5.3\% of 
flux from the quasar in the D$_{A}$ wavelength range,
and proves that the power-law approximation is inadequate to
perform a D$_{A}$ measurement that attempts 1\% accuracy
\citep{tytler04c,jena04a}.
The distribution of $\delta F$ is shown in Figure \ref{fig:p12},
and it is not a Gaussian. 
The PDF of $\delta F$ is asymmetric and has long tail toward 
small $\delta F$ value.

\outline{draft}{$\delta F$ and the origin}
There are two major reasons for the missing flux.
The first reason is that the D$_{A}$ wavelength range is still 
in the tail of the prominent Ly$\beta$/O~VI and Ly$\alpha$
emission lines.
For example, the blueward tail of the Ly$\alpha$ emission starts 
near $\lambda$1160 which is 10\AA\ below the $\lambda$1170
upper limit of D$_{A}$ wavelength range.
The effect from the tails of Ly$\beta$ and Ly$\alpha$
emission is common for all of the quasars as we can see in 
Figure \ref{fig:p7}-\ref{fig:p11}. 
This contribution is about 4-5\%, and we always miss this fraction 
of the flux, meaning $\delta F$ is always less than unity.

\outline{draft}{Second Source for $\delta F$}
The second reason is the contribution from the weak 
emission lines in the Ly$\alpha$ forest.
The continuum level at the emission lines is naturally 
above the power-law extrapolation,
therefore, we would always expect to miss flux from the emission
lines, making $\delta F$ always less than unity.
The low ionization emission lines in the Ly$\alpha$ forest are prominent
for {\it Class III} \& {\it Class IV} quasars.
Four quasars in {\it Class III} \& {\it Class IV} have $\delta F$ less
than 0.9, meaning we miss more than 10\% of the flux if we use
power-law extrapolation.
The quasar which has the smallest $\delta F$, 0.86, is Q1544+4855.
This quasar is shown on the top panel of Figure \ref{fig:p11}
and the power-law continuum fit looks sensible.
Together with the tails of Ly$\beta$ and Ly$\alpha$ emissions,
the low ionization emission lines in the Ly$\alpha$ forest (Fe~II, Fe~III) 
contribute 14\% of flux which is significant and should not be neglected.

\outline{draft}{Discussion}
We expect that we need to apply at least a $\delta F$ = 0.947 correction 
to the D$_{A}$ measurements in the literature.
The discrepancy between the past D$_{A}$ measurements and
\citet{bernardi03} using the weak emission profile fitting method is shown 
in \citet[][Figure 22]{tytler04c}.  
The disagreement is approximately 5\% at redshift $2 < z < 3$, 
and in terms of the mean flux, the power-law fitted values 
\citep{press93a,steidel87a} are always 
above that of profile fitted ones \citep{bernardi03}.
The correction factor, $\delta F$ = 0.947, explains 
this disagreement well.

\outline{draft}{dF evolution with redshift}
However, we expect that $\delta F$ changes with redshift, and it is
crucial to include the effect from the weak emission lines to investigate
the reionization epoch.
Known as the Baldwin effect \citep{baldwin77}, the emission profile of
lines, such as C~IV, and the luminosity of the quasar are correlated.
Because of the anticorrelation between equivalent width of C~IV 
and the luminosity, we expect that {\it Class III} quasars to be the
brightest since they have the smallest C~IV equivalent width.
Due to the selection effect, we would expect to observe 
the brightest quasars at high redshifts.
Therefore, the fraction of classes for the observed quasars
should change with redshift.
In fact, the highest redshift quasars reported by 
\citet{fan01}, \citet{becker01} and \citet{djorgovski01} probably 
belong to {\it Class III},
because they all show weak Si~II $\lambda 1304$ emission line implying 
they have weak emission lines in the Ly$\alpha$ forest regain.
J104433.04-012502.2 (z=5.80) and J08643.85+005453.3 (z=5.82) 
definitely belong to {\it Class III} since they have broad Ly$\alpha$, 
N~V, and Si~II emission lines and low Ly$\alpha$/N~V emission 
intensity ratio.
The mean flux correction $\delta F$ of {\it Class III} quasars are 
in the range of 0.91-0.97, 
meaning the power-law extrapolation misses 3-9\% of the flux for 
{\it Class III} quasars. 
We note that the reported 1-$\sigma$ error of the residual mean flux 
at redshift z=5.75 by \citet{becker01} is 0.03.
The contribution from $\delta F$ is bigger than their claimed error,
and it is systematic.
This correction would bring the observed mean flux down by 3-9\%
percent and would bring D$_{A}$ up by the same fraction.
Therefore, it is crucial to take into account the $\delta F$ correction 
in order to investigate the reionization epoch.



\section{Summary \NOTE{section h} } \label{sec:h}
\outline{draft}{Paper Achievement}
We analyzed the wide variety of the emission line profiles 
in the Ly$\alpha$ forest both in a quantitative and qualitative way.
We used PCA to describe the variety of quasar spectra, and
we found that 1161 pixels of data ($\lambda\lambda1020-1600$ with 0.5 \AA\
binning) can be summarized by the primary seven PCS coefficients because
the pixels are not independent but are strongly correlated with each other.
We presented, for the first time,
the idea of generating artificial quasar spectra.
Our artificial quasar spectra should be useful in testing the detections,
in calibrations, and in simulations.
We introduced five classes to differentiate the families of quasar 
spectra, and showed how the classification
can guide us to find the continuum level in the Ly$\alpha$ forest. 
It is essential to account for emission line features in
the Ly$\alpha$ forest to perform a precise measurement
of the mean flux in order to probe cosmological parameters,
the UV background, and the reionization epoch,
otherwise, on average, the commonly used power-law extrapolation continuum 
misses 5.3\% of the flux, and we have cases when it misses up to 14\% 
of the flux.

\outline{draft}{}
To investigate the high redshift Ly$\alpha$ forest, we showed the need to
account for the emission lines in the Ly$\alpha$ forest.
An emission line profile continuum fitting method by 
\citet{bernardi03}, or improvement of the PCA method in Paper I would
be useful for a large data set such as the Sloan Digital Sky Survey.
If we can study the redshift evolution of the quasar spectra and 
if we can estimate the constituents of classes at a certain redshift,
we would be able to estimate the mean flux statistically using
the mean flux correction factor $\delta F$.
For precision cosmology, the formalized we presented here should
play an important role.

I thank Regina Jorgenson, Kim Griest, Jeff Cooke, Carl Melis,
Tridi Jena and Geffrey So for their careful reading of this manuscript 
and encouragnements.
I thank David Tytler and David Kirkman for the comments 
on the manuscript and the discussions on the Ly$\alpha$ forest studies.
I thank Art Wolfe and Chris Hawk for the discussions on the emission line 
identifications.
I am grateful to Paul Francis who provided the code and LBQS spectra.
Paul Hewett kindly sent the error arrays for those spectra.
Wei Zheng and Buell Januzzi kindly provided 
copies of HST QSO spectra that we used before we located the invaluable
collection of HST spectra posted to the web by Jill Bechtold.
I thank the Okamura Group at the University of Tokyo from whom I learned 
the basics of PCA.  
This work was supported in part by NASA grants NAG5-13113 and
HST-AR-10288.01-A from the STScI and by NSF grant  AST-0098731.

\appendix
\section{Appendix : Weak Emission Lines 
in the Ly$\alpha$ Forest  \NOTE{section c} } 
\label{sec:c}
\outline{draft}{3 Weak Emission Line intro}
It has been suggested that there exist weak emission lines in the
Ly$\alpha$ forest \citep{zheng97,telfer02,bernardi03,scott04a},
however, their identities, strengths, and line profiles are not 
well understood.
More importantly, we wish to know how they are correlated with other
emission lines so that we can predict the strength and profiles of the
emission lines in the Ly$\alpha$ forest.
We attempt to find the identities of the three weak emission lines 
in the Ly$\alpha$ forest reported by \citet{tytler04a}. 
In this paper, we use their measured wavelengths :
$\lambda=1070.95, \lambda=1123.17, \lambda=1175.88$ \AA .


\subsection{Line at $\lambda=1070.95$ \AA\ : Fe~II}
\outline{draft}{ID in the literature is wrong}
\citet{zheng97} and \citet{vandenberk01} identified 
this line as Ar~I $\lambda$1066.66,
but the contribution from Ar~I cannot be this large. 
The Ar~I line has another transition at $\lambda$1048.22 whose transition
probability ($A_{ik}$ =  $4.94 \times 10^{8} s^{-1}$) is
stronger than that of $\lambda$1066.66 ($A_{ik}= 1.30 \times 10^{8} s^{-1}$).
However, there is no clear sign of the $\lambda$1048.22 line feature in
50 {\it HST} spectra or 79 quasar spectra in \citet{tytler04c}.
In addition, 
the $\lambda1070$ line has a broad and asymmetric profile which suggests 
that this line is a blend of multiple lines.
Thus, $\lambda1070$ is not likely to be a single Ar~I line.

\outline{draft}{Telfer \& Scott are wrong}
\citet{telfer02} labeled this $\lambda1070$ as N~II + He~II line,
and mentioned S~IV as a possible candidate.
\citet{scott04a} identify four lines in the proximity of this wavelength
: S~IV doublet ($\lambda\lambda$1062, 1073) and N~II+He~II+Ar~I 
($\lambda$1084).
As we have seen in \S \ref{sec:i}, the $\lambda1070$ line 
correlates with low ionization lines such as Si~II and C~II.  
S~IV does not fit into this category.
N~II seems to be a good candidate, but the N~II lines peaks around 1085 \AA\
which is 15 \AA\ away from what we see.
It is unlikely that we have 15 \AA\ of wavelength error.
For the same reason, He~II (1085\AA), a high ionization line, 
is not likely to be the dominant contributor. 
However, it is reasonable to expect to have an He~II line since
He~II $\lambda$1640 is often seen in quasar spectra.
This $\lambda1070$ line is seen in the Q1009+2956 and Q1243+3047 spectra 
for which we have high S/N Keck HIRES spectra with a FWHM=0.0285 \AA\ 
resolution at this wavelength \citep{burles98b,kirkman03a}.
If $\lambda1070$ is comprised of the lines suggested by \citet{scott04a}, 
we would be able to resolve the individual lines.
However, none of the individual emission line are resolved.
This fact implies that this emission line is comprised of numerous weak
lines and that they are probably low ionization lines because of the good
correlation with other low ionization lines.

\outline{draft}{Could be FeII but ..}
Fe~II suits such a description, and in fact, Fe~II has a series of UV
transitions around this wavelength range: $\lambda\lambda$1060-1080.
However, we don't see other expected Fe~II emission lines. 
If this $\lambda1070$ is Fe~II, we would expect to see the 
Fe~II UV10 multiplet around $\lambda$1144 , which is 
supposed to be stronger than the $\lambda1070$ line. 
But no sign of an emission line is seen at that wavelength.
Therefore, Fe~II identification may be wrong or there may be 
a mechanism that we are not aware of that prevents the
expected $\lambda$1144 emission line.  

\subsection{$\lambda=1123.17$ \AA\ : Fe~II + Fe~III}
\outline{draft}{FeIII}
\citet{telfer02,vandenberk01} identify $\lambda1123$ as Fe~III.
The f-value weighted Fe~III UV1 multiplet has a wavelength at 1126.39 \AA\
which is very close to what is observed by \citet{tytler04c}.
The distribution of the multiplet lines 
matches the broad $\lambda1123$ feature found in quasars.
There are no other major resonance lines in this wavelength range except C~I.
In the wavelength range $\lambda\lambda1114-1200$, C~I has a series of 
lines,
and there exists stronger lines redward of the Ly$\alpha$ 
emission line: $\lambda\lambda1115-1193, \lambda\lambda1277-1280$.
However, we do not see these redward C~I emission lines, and
there is no sign of correlation between $\lambda1123$ and
these possible lines (Paper~I).
Thus, $\lambda1123$ is probably Fe~III.
In addition to Fe~III, Fe~II also has UV11-14 multiplets around this
wavelength: $\lambda\lambda1121 - 1133$. 
Given the fact that this $\lambda1123$ line is well correlated with 
$\lambda1070$ line, it is reasonable to expect to see Fe~II 
lines here as well, if the identification of $\lambda1070$ is Fe~II.

\subsection{$\lambda=1175.88$ \AA\ : CIII$^{*}$}
\outline{draft}{CIII}
\citet{telfer02,vandenberk01} identified $\lambda1175$ as C~III$^{*}$,
although as shown in Table \ref{tbl:e}, 
we do not have a clear detection of this line
because of it being a weak and narrow feature.
We can see the $\lambda1175$ line in the {\it HST} composite spectrum 
\citep{telfer02}, the SDSS composite spectrum \citep{vandenberk01}, and 
in 11 out of 79 quasar spectra in \citet{tytler04c}.
Therefore, we are confident that this $\lambda1175$ is a real 
emission line.
The $f-$value weighted wavelength of the C~III$^{*}$ line 
is $\lambda$1175.5289 which matches well the observed wavelength. 
There is no major resonance line at this wavelength.


\subsection{Other Possible Emission Lines in the Ly$\alpha$ Forest}
\outline{draft}{SiII}
\citet{tytler04a} and \citet{telfer02} reported observing the Si~II 
$\lambda1195$ line in their spectra.
We do not have any clear detection of the Si~II $\lambda$1195 line
in the 50 quasar spectra.
Since other Si~II lines, $\lambda\lambda$1265, 1304, are clearly 
seen 
redward of the Ly$\alpha$ emission line, it is plausible to
expect the Si~II $\lambda$1195 line.

\outline{draft}{SiIII}
Si~III has a transition at $\lambda$1206.50.  
Since we see both Si~II and Si~IV 
redward of Ly$\alpha$ emission, it is natural to expect to see Si~III.
However, no detection is reported in the literature, and we do not have
any positive detection in the 50 quasar spectra.
The Si~III $\lambda1206$ emission line is probably too weak or possibly 
overwhelmed by broad Ly$\alpha$ emission, 
only 10 \AA\ away,
whose equivalent width is 
often greater than 100\AA .

\begin{deluxetable}{lrrrrrrrrrr}
\tablewidth{0pt}
\tabletypesize{\scriptsize}
\tablecaption{Eigenvalue and Residual Variance Fraction \label{tbl:a}}

\tablehead{
\colhead{Component : $j$} & 
\colhead{1}  &
\colhead{2}  &
\colhead{3}  &
\colhead{4}  &
\colhead{5}  &
\colhead{6}  &
\colhead{7}  &
\colhead{8}  &
\colhead{9}  &
\colhead{10} 
}

\startdata
Eigenvalue $\lambda_{j}^{2}$ & 57.204 & 12.985 & 5.528 & 4.614 & 2.516 & 2.189 & 1.293 & 0.605 & 0.541 & 0.453\\
$\lambda_{j}$   & 7.563 & 3.604 & 2.351 & 2.148 & 1.586 & 1.479 & 1.137 & 0.778 & 0.735 & 0.673\\
Residual Variance Fraction f($j$) & 0.637 & 0.145 & 0.062 & 0.051 & 0.028 & 0.024 & 0.014 & 0.007 & 0.006 & 0.005\\
Cummulative Residual Variance Fraction & 0.637 & 0.781 & 0.843 & 0.894 & 0.922 & 0.947 & 0.961 & 0.968 & 0.974 & 0.979
\enddata
\tablecomments{Cummulative residual variance fracion is a simple sum of
residual variance fraction upto $j$th PCS.}
\end{deluxetable}

\begin{deluxetable}{rlrrrrrrrrrrrrr}
\tabletypesize{\tiny}
\tablecaption{The mean flux correction factor $\delta F$ and 
equivalent width of emission lines\label{tbl:e} \NOTE{tbl:e}}
\tablewidth{0pt}
\tablehead{
\colhead{ $i$  } &
\colhead{Quasar} &
\colhead{$\delta$F} &
\colhead{Ly$\beta$/OVI} &
\colhead{Fe~II} &
\colhead{Fe~III} &
\colhead{C~III$^{*}$} &
\colhead{Ly$\alpha$} &
\colhead{N~V} &
\colhead{Si~II} &
\colhead{Si~II} &
\colhead{C~II} &
\colhead{Si~IV} &
\colhead{C~IV} \\

\colhead{} &
\colhead{} &
\colhead{} &
\colhead{1025} &
\colhead{1071} &
\colhead{1123} &
\colhead{1176} &
\colhead{1216} &
\colhead{1240} &
\colhead{1263} &
\colhead{1307} &
\colhead{1335} &
\colhead{1397} &
\colhead{1549} \\

\colhead{} &
\colhead{} &
\colhead{} &
\colhead{(\AA)} &
\colhead{(\AA)} &
\colhead{(\AA)} &
\colhead{(\AA)} &
\colhead{(\AA)} &
\colhead{(\AA)} &
\colhead{(\AA)} &
\colhead{(\AA)} &
\colhead{(\AA)} &
\colhead{(\AA)} &
\colhead{(\AA)} 
}

\startdata
&{\bf Class Zero}&&&&&&&&&&&&&\\
\tableline
13 & Q0947+3940 & 0.972 & 8.8 & 1.4 & 0.0 & 0.0 & 102.9 & 0.7 & 0.5 & 0.0 & 0.0 & 13.2 & 59.8\\
25 & Q1229-0207 & 0.953 & 9.5 & 0.9 & 0.2 & 0.0 & 87.7 & 0.4 & 0.1 & 1.8 & 0.0 & 6.9 & 47.4\\
34 & Q1424-1150 & 0.903 & 9.5 & 1.1 & 0.0 & 0.0 & 131.9 & 0.6 & 0.0 & 0.0 & 0.0 & 8.1 & 50.4\\
37 & Q1538+4745 & 0.926 & 8.2 & 0.4 & 0.9 & 0.0 & 132.9 & 0.0 & 0.0 & 0.4 & 0.0 & 6.8 & 43.4\\
48 & Q2340-0339 & 0.963 & 5.1 & 0.3 & 0.0 & 0.0 & 83.5 & 0.7 & 0.0 & 0.0 & 0.0 & 5.7 & 41.5\\
\tableline
  & Average & 0.944 & 8.2 & 0.8 & 0.2 & 0.0 & 107.8 & 0.5 & 0.1 & 0.4 & 0.0 & 8.1 & 48.5\\
  & STD & 0.028 & 1.8 & 0.5 & 0.4 & 0.0 & 23.6 & 0.3 & 0.2 & 0.8 & 0.0 & 2.9 & 7.2\\
\tableline
&{\bf Class I}&&&&&&&&&&&&&\\
\tableline
1 & Q0003+1553 & 0.940 & 8.6 & 0.1 & 0.0 & 0.0 & 115.7 & 0.0 & 0.0 & 2.2 & 0.0 & 6.5 & 55.2\\
8 & Q0439-4319 & 0.970 & 13.3 & 0.0 & 0.0 & 0.0 & 119.2 & 1.1 & 0.0 & 1.2 & 0.0 & 7.3 & 58.0\\
11 & Q0637-7513 & 0.980 & 15.6 & 2.5 & 0.0 & 0.0 & 87.2 & 2.2 & 0.3 & 0.0 & 0.0 & 8.6 & 38.9\\
14 & Q0953+4129 & 0.937 & 12.1 & 1.9 & 0.3 & 0.0 & 124.2 & 1.4 & 0.3 & 0.0 & 0.0 & 10.6 & 63.6\\
18 & Q1007+4147 & 0.943 & 11.6 & 2.0 & 0.1 & 0.1 & 145.4 & 1.9 & 0.0 & 2.0 & 0.3 & 10.4 & 65.5\\
22 & Q1137+6604 & 0.939 & 10.8 & 1.8 & 0.2 & 0.0 & 99.2 & 0.6 & 0.2 & 0.6 & 0.0 & 7.3 & 45.1\\
24 & Q1216+0655 & 0.959 & 9.9 & 0.9 & 0.0 & 0.0 & 124.7 & 0.0 & 0.0 & 0.6 & 0.0 & 8.8 & 52.4\\
32 & Q1354+1933 & 0.974 & 10.3 & 0.2 & 0.1 & 0.1 & 134.6 & 0.5 & 0.5 & 2.0 & 1.1 & 7.9 & 65.5\\
39 & Q1622+2352 & 0.971 & 9.3 & 0.7 & 0.0 & 0.1 & 111.3 & 0.2 & 0.0 & 0.0 & 0.0 & 7.7 & 65.2\\
41 & Q1821+6419 & 0.986 & 13.6 & 0.3 & 0.0 & 0.0 & 115.3 & 0.2 & 0.0 & 0.0 & 0.0 & 7.8 & 50.6\\
42 & Q1928+7351 & 0.982 & 10.2 & 0.0 & 0.0 & 0.0 & 149.4 & 1.9 & 0.0 & 0.0 & 0.0 & 12.4 & 74.3\\
45 & Q2243-1222 & 0.966 & 8.1 & 0.3 & 0.0 & 0.1 & 103.2 & 0.0 & 0.0 & 2.2 & 0.2 & 8.1 & 51.6\\
49 & Q2344+0914 & 0.963 & 13.9 & 0.8 & 0.0 & 0.0 & 143.1 & 1.8 & 0.5 & 1.6 & 0.0 & 5.3 & 50.9\\
\tableline
  & Average & 0.962 & 11.3 & 0.9 & 0.1 & 0.0 & 121.0 & 0.9 & 0.1 & 1.0 & 0.1 & 8.4 & 56.7\\
  & STD & 0.017 & 2.3 & 0.9 & 0.1 & 0.0 & 18.7 & 0.8 & 0.2 & 0.9 & 0.3 & 1.9 & 9.8\\
\tableline
&{\bf Class II}&&&&&&&&&&&&&\\
\tableline
2 & Q0026+1259 & 0.950 & 9.6 & 2.4 & 0.1 & 0.0 & 92.1 & 3.6 & 0.1 & 0.0 & 0.0 & 9.2 & 19.4\\
4 & Q0159-1147 & 0.948 & 2.8 & 0.3 & 0.9 & 0.0 & 52.8 & 1.0 & 0.0 & 1.1 & 0.8 & 4.2 & 16.8\\
5 & Q0349-1438 & 0.974 & 1.8 & 0.0 & 0.0 & 0.1 & 54.9 & 0.0 & 0.0 & 0.0 & 0.0 & 4.0 & 28.0\\
6 & Q0405-1219 & 0.950 & 6.9 & 0.7 & 1.1 & 0.0 & 92.5 & 0.4 & 0.0 & 1.1 & 0.0 & 6.5 & 35.0\\
7 & Q0414-0601 & 0.958 & 6.1 & 0.0 & 0.1 & 0.0 & 114.0 & 0.9 & 0.0 & 0.0 & 0.0 & 5.1 & 40.1\\
9 & Q0454-2203 & 0.935 & 11.6 & 4.2 & 0.0 & 0.0 & 105.7 & 2.2 & 0.3 & 1.1 & 0.0 & 7.2 & 33.9\\
12 & Q0923+3915 & 0.973 & 7.2 & 0.0 & 0.1 & 0.0 & 95.1 & 0.1 & 0.0 & 1.2 & 0.0 & 5.1 & 46.4\\
15 & Q0954+5537 & 0.983 & 3.3 & 0.0 & 0.4 & 0.1 & 42.6 & 0.3 & 1.1 & 0.0 & 0.0 & 1.9 & 17.9\\
20 & Q1104+1644 & 0.958 & 5.7 & 0.7 & 0.0 & 0.0 & 104.5 & 0.0 & 0.0 & 0.5 & 0.0 & 7.8 & 54.1\\
27 & Q1252+1157 & 0.939 & 6.1 & 2.7 & 2.4 & 0.0 & 72.7 & 1.0 & 0.2 & 0.0 & 0.5 & 6.0 & 25.5\\
40 & Q1637+5726 & 0.945 & 4.5 & 0.9 & 0.2 & 0.0 & 77.6 & 0.7 & 0.0 & 0.3 & 0.0 & 5.4 & 33.2\\
47 & Q2251+1552 & 1.015 & 7.1 & 2.3 & 0.0 & 0.0 & 60.1 & 1.0 & 0.0 & 0.0 & 0.0 & 2.6 & 23.2\\
\tableline
  & Average & 0.960 & 6.1 & 1.2 & 0.4 & 0.0 & 80.4 & 0.9 & 0.1 & 0.4 & 0.1 & 5.4 & 31.1\\
  & STD & 0.022 & 2.8 & 1.4 & 0.7 & 0.0 & 23.7 & 1.0 & 0.3 & 0.5 & 0.2 & 2.1 & 11.6\\
\tableline
&{\bf Class III}&&&&&&&&&&&&&\\
\tableline
16 & Q0959+6827 & 0.865 & 3.6 & 4.0 & 2.9 & 0.1 & 81.0 & 3.1 & 0.7 & 1.3 & 1.3 & 12.1 & 25.2\\
17 & Q1001+2910 & 0.949 & 3.4 & 3.4 & 2.4 & 0.0 & 68.7 & 2.0 & 0.7 & 2.9 & 2.1 & 9.3 & 26.2\\
21 & Q1115+4042 & 0.957 & 7.3 & 4.3 & 3.2 & 0.0 & 82.3 & 1.8 & 0.3 & 2.5 & 1.5 & 8.6 & 29.7\\
23 & Q1148+5454 & 0.928 & 3.4 & 3.9 & 2.8 & 0.0 & 108.4 & 1.1 & 0.9 & 1.8 & 1.5 & 14.1 & 30.8\\
26 & Q1248+4007 & 0.937 & 2.9 & 3.0 & 3.2 & 0.1 & 111.6 & 2.1 & 0.6 & 1.6 & 0.6 & 12.2 & 28.6\\
28 & Q1259+5918 & 0.954 & 2.8 & 2.5 & 1.8 & 0.0 & 63.5 & 1.0 & 1.7 & 2.7 & 0.7 & 10.1 & 19.8\\
29 & Q1317+2743 & 0.963 & 3.2 & 0.8 & 1.3 & 0.0 & 57.2 & 1.5 & 0.4 & 0.5 & 0.1 & 7.1 & 17.9\\
30 & Q1320+2925 & 0.977 & 1.1 & 3.9 & 1.2 & 0.0 & 58.6 & 1.0 & 0.6 & 0.4 & 0.1 & 5.7 & 14.4\\
36 & Q1444+4047 & 0.925 & 4.0 & 3.0 & 3.8 & 0.1 & 65.0 & 3.6 & 0.0 & 0.3 & 0.7 & 9.9 & 24.0\\
43 & Q2145+0643 & 0.943 & 1.6 & 0.0 & 0.0 & 0.0 & 71.3 & 0.6 & 0.0 & 0.0 & 0.3 & 1.6 & 35.9\\
44 & Q2201+3131 & 0.928 & 4.8 & 0.6 & 1.2 & 0.0 & 69.9 & 0.0 & 0.0 & 0.0 & 0.0 & 7.6 & 28.8\\
\tableline
  & Average & 0.939 & 3.5 & 2.7 & 2.2 & 0.0 & 76.1 & 1.6 & 0.5 & 1.3 & 0.8 & 8.9 & 25.6\\
  & STD & 0.029 & 1.6 & 1.5 & 1.1 & 0.0 & 18.5 & 1.1 & 0.5 & 1.1 & 0.7 & 3.5 & 6.2\\
\tableline
&{\bf Class IV}&&&&&&&&&&&&&\\
\tableline
3 & Q0044+0303 & 0.965 & 8.6 & 1.3 & 0.0 & 0.0 & 146.0 & 1.5 & 0.1 & 1.7 & 0.1 & 11.3 & 71.2\\
10 & Q0624+6907 & 0.888 & 8.2 & 5.1 & 3.0 & 0.0 & 174.5 & 3.9 & 0.6 & 2.5 & 1.3 & 11.3 & 51.7\\
19 & Q1100+7715 & 0.948 & 9.7 & 1.7 & 0.0 & 0.0 & 101.8 & 1.0 & 0.0 & 0.0 & 0.0 & 8.5 & 72.1\\
31 & Q1322+6557 & 0.925 & 8.3 & 1.8 & 1.7 & 0.0 & 138.5 & 2.7 & 0.0 & 3.3 & 0.8 & 9.3 & 55.5\\
33 & Q1402+2609 & 0.865 & 6.3 & 3.1 & 4.7 & 0.0 & 82.0 & 1.8 & 0.8 & 2.3 & 1.8 & 5.1 & 31.7\\
35 & Q1427+4800 & 0.944 & 13.9 & 1.8 & 0.1 & 0.0 & 96.9 & 0.7 & 0.0 & 0.6 & 0.2 & 16.9 & 57.3\\
38 & Q1544+4855 & 0.860 & 2.9 & 5.7 & 5.5 & 0.0 & 119.5 & 1.0 & 1.6 & 5.1 & 3.6 & 16.5 & 33.5\\
46 & Q2251+1120 & 0.937 & 10.5 & 0.8 & 0.0 & 0.1 & 141.6 & 4.5 & 0.0 & 2.1 & 0.0 & 7.8 & 69.8\\
50 & Q2352-3414 & 0.937 & 9.0 & 0.8 & 1.3 & 0.0 & 129.4 & 0.4 & 0.0 & 0.7 & 0.2 & 6.6 & 61.8\\
\tableline
  & Average & 0.919 & 8.6 & 2.4 & 1.8 & 0.0 & 125.6 & 1.9 & 0.3 & 2.0 & 0.9 & 10.4 & 56.1\\
  & STD & 0.038 & 3.0 & 1.8 & 2.1 & 0.0 & 28.7 & 1.5 & 0.6 & 1.6 & 1.2 & 4.1 & 15.1\\
&{\bf Total}&  &&&&&&&&&&&&\\
\tableline
  & Average & 0.947 & 7.5 & 1.6 & 0.9 & 0.0 & 100.9 & 1.2 & 0.3 & 1.0 & 0.4 & 8.1 & 42.8\\
  & STD & 0.031 & 3.7 & 1.5 & 1.4 & 0.0 & 30.4 & 1.1 & 0.4 & 1.1 & 0.7 & 3.3 & 17.0\\
\enddata


\end{deluxetable}
\end{document}